\documentclass[sigconf]{acmart}

\usepackage{booktabs}
\usepackage{enumerate}
\usepackage{epsfig}
\usepackage{graphicx}
\usepackage{epstopdf}
\usepackage{amsmath}
\usepackage[linesnumbered,ruled]{algorithm2e}
\usepackage{url}
\usepackage{listings}
\usepackage{subfigure}
\usepackage{rotating}
\usepackage{lscape}
\usepackage{verbatim}
\usepackage{hhline}
\usepackage{textcomp,booktabs}
\usepackage{xcolor}
\usepackage{paralist}
\usepackage{hyperref}
\usepackage{grffile}
\usepackage{mathtools}
\usepackage{amsthm}
\usepackage{hyphenat}
\usepackage{datatool}
\usepackage{numprint}
\npthousandsep{,}
\usepackage{color, colortbl}
\usepackage{flushend}

\DTLsetseparator{,}

\DTLloaddb{bbcdata}{./stats.txt}

\DTLforeach*
{bbcdata}% database label
{\totalapps=total_apps, \numreviews=collected_reviews,\numexchangers=num_exchangers,\numregular=num_regular,\numworkers=num_workers,\numslowsnaps=num_slow_snaps,\numfastsnaps=num_fast_snaps}

\newcommand{\newmaterial}[1]{\textcolor{black}{{#1}}} 
\copyrightyear{2021}
\acmYear{2021}
\setcopyright{acmcopyright}\acmConference[IMC '21]{ACM Internet Measurement Conference}{November 2--4, 2021}{Virtual Event, USA}
\acmBooktitle{ACM Internet Measurement Conference (IMC '21), November 2--4, 2021, Virtual Event, USA}
\acmPrice{15.00}
\acmDOI{10.1145/3487552.3487837}
\acmISBN{978-1-4503-9129-0/21/11}

\begin{document}
\begin{filecontents*}{./stats.txt}
total_apps,collected_reviews,num_exchangers,num_regular,num_workers,num_fast_snaps,num_slow_snaps
12341,110511637,31,223,549,57770204,592045
\end{filecontents*}

\title[RacketStore: Measurements of ASO Deception]{RacketStore: Measurements of ASO Deception in Google Play via Mobile and App Usage}

\author{Nestor Hernandez}
\affiliation{
	\institution{FIU, Miami, USA}
}
\email{nestorghh@gmail.com}

\author{Ruben Recabarren}
\affiliation{
	\institution{FIU, Miami, USA}
}
\email{recabarren@gmail.com}
\author{Bogdan Carbunar}
\affiliation{
	\institution{FIU, Miami, USA}
}
\email{carbunar@cs.fiu.edu}
\author{Syed Ishtiaque Ahmed}
\affiliation{
	\institution{University of Toronto, Toronto, CA}
}
\email{ishtiaque@cs.toronto.edu}

\renewcommand{\shortauthors}{N. Hernandez, R. Recabarren, S.I. Ahmed, B. Carbunar}

\begin{abstract}
Online app search optimization (ASO) platforms that provide bulk installs and fake reviews for paying app developers in order to fraudulently boost their search rank in app stores, were shown to employ diverse and complex strategies that successfully evade state-of-the-art detection methods. In this paper we introduce RacketStore, a platform to collect data from Android devices of participating ASO providers and regular users, on their interactions with apps which they install from the Google Play Store. We present measurements from a study of 943 installs of RacketStore on 803 unique devices controlled by ASO providers and regular users, that consists of 58,362,249 data snapshots collected from these devices, the 12,341 apps installed on them and their 110,511,637 Google Play reviews. We reveal significant differences between ASO providers and regular users in terms of the number and types of user accounts registered on their devices, the number of apps they review, and the intervals between the installation times of apps and their review times.
We leverage these insights to introduce features that model the usage of apps and devices, and show that they can train supervised learning algorithms to detect paid app installs and fake reviews with an F1-measure of 99.72\% (AUC above 0.99), and detect devices controlled by ASO providers with an F1-measure of 95.29\% (AUC = 0.95). We discuss the costs associated with evading detection by our classifiers and also the potential for app stores to use our approach to detect ASO work with privacy.
\end{abstract}




\maketitle

\section{Introduction}
\label{sec:introduction}

The global mobile application market is worth hundreds of billions of USD and is expected to grow by more than 10\% per year until 2027~\cite{AppMarketShare}. To stand out among millions of apps hosted in app stores~\cite{AppCountGooglePlay, AppCountAppleStore} and get a share of this market, many app developers resort to app search optimization (ASO) to increase the rank of their apps during search. ASO platforms use a variety of techniques to achieve this~\cite{Apptentive}, including providing retention installs~\cite{RetentionInstall} and posting fake reviews~\cite{MobiASO, Boostyourapps}. Such activities can be illegal in countries like the US~\cite{15USCode45}, Canada~\cite{74.02Canada}, Australia~\cite{T19}, are banned in the EU~\cite{0529EU}, violate the terms of service of app stores~\cite{GooglePlayToS, AppleStoreToS}, and influence users to install and purchase low quality apps and even malware~\cite{RRCC17,WLLVGLTCX18,HWHLTCGWX20}.

Identifying ASO-promoted apps and the accounts from which they are promoted allows app stores to filter fake reviews and ratings, generate more accurate install count and aggregate rating values thus compute more accurate search ranks for apps, and enable users to make better informed app-installation decisions.

A key to achieve this is to build an accurate understanding of the behaviors and strategies employed by fraudulent app search optimization (ASO) workers. In previous work, Farooqi et al.~\cite{FFLMSV20} have shown that incentivized app install platforms (IIP) are able to provide thousands of installs that successfully evade Google defenses. Rahman et al.~\cite{RHRAC19} have reported a variety of detection-avoidance techniques employed by organizations that specialize in retention installs and fake reviews. Such techniques include crowdsourcing ASO work to {\it organic workers}, who use their personal devices to conceal ASO work among everyday activities.

Identifying organic ASO activities is an open problem due to the ability of such workers to evade existing detection solutions, e.g., that leverage lockstep behaviors~\cite{CYYP14, XZ14, SMJEKV15, TZXZZ15, YKA16, XZ15.SIAM, SLK15, XZ15.WiSec, LFWMS17, YMGYLSWL19,HRRC18} or review bursts~\cite{NAGKV19, XWLY12, FMLHCG13, HTS16, LFWMS17, HSBGAKMF16, SLLTZ16, XZLW16, X13, GGF14, BXGPF13, LNJLL10, MVLG13, MKLWHCG13, KCS18, LCNK17, YKA16, FLCS15, DFJKS14, LNJLL10, RHCC18, RHCC20}.

\begin{figure}
\includegraphics[width=0.99\columnwidth]{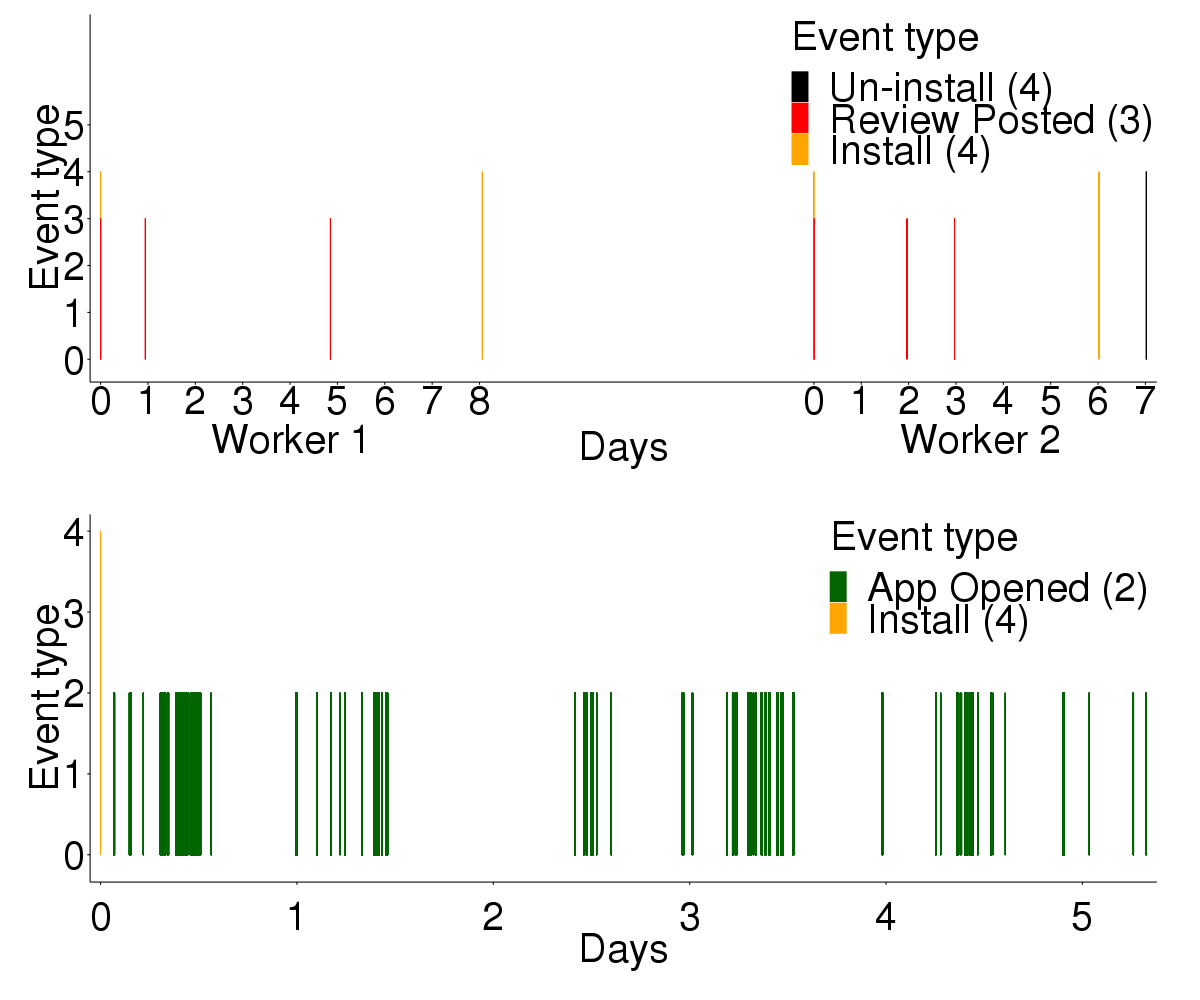}
\vspace{-15pt}
\caption{On-device app interaction timelines for two ASO workers (top) and one regular user (bottom). Worker timelines start with the app installation event (type 4 on y axis), followed by several review posting events across several days (type 3), with no interaction with the app. In contrast, the regular user timeline shows frequent interaction with the app, e.g., placing the app in the foreground (type 2 event), but no review even after 5 days of monitoring.}
\label{fig:combined_timeline}
\vspace{-20pt}
\end{figure}

In an effort to determine whether solutions can be developed to detect these ASO strategies, in this paper we seek to measure and compare the device and app usage of ASO workers and regular users. Our work is partially motivated by Farooqi et al.~\cite{FFLMSV20}'s finding that ASO workers lack interest in the apps that they promote. We conjecture this also results in, e.g., workers posting reviews for promoted apps soon after installing them, see Figure~\ref{fig:combined_timeline}.

To enable such measurements, we develop RacketStore, a platform to collect and analyze app and device-use data from consenting ASO workers and regular users. The RacketStore mobile app periodically collects data from the devices where it is installed, e.g., the foreground app and the list of installed apps with 5s granularity, and the types and number of registered accounts, with 2 min granularity. The RacketStore server aggregates this information with data collected from the Play Store and VirusTotal~\cite{VirusTotal}. \newmaterial{RacketStore first discloses the types of data it collects. RacketStore collects the data only after receiving participant consent} ($\S$~\ref{sec:methods:ethical} and Appendix~\ref{appendix:ethics}).

We present measurements from a study of ASO workers and regular users recruited to keep the RacketStore app installed on their devices for at least two days. In total, RacketStore was installed 943 times on \FPeval{\result}{clip(\numregular+\numworkers+\numexchangers)} \numprint{\result} unique devices: \FPeval{\aresult}{clip(\numworkers+\numexchangers)} \numprint{\aresult} devices controlled by ASO workers recruited from Facebook groups that specialize in ASO work, and \numregular~ devices of regular Android users recruited through ads purchased in Instagram. We have collected \FPeval{\result}{clip(\numslowsnaps+\numfastsnaps)} \numprint{\result} snapshots from the participating devices, including their \numprint{\totalapps} apps installed and in-use, and their \numprint{\numreviews} reviews from the Play Store.

We found that ASO work continues to be successful and evade app store detection: The worker-controlled devices of participants in our studies had 10,310 Gmail accounts registered on them and at the time of writing, Google Play was still displaying 217,041 reviews posted from them.

Measurements reveal that many participant ASO worker devices have organic-indicative behaviors, i.e., similar to those of regular devices, in terms of their app churn (daily installed and uninstalled apps), permissions granted, the total number of installed apps, stopped apps, or daily used apps. However, we found significant differences between regular user and worker-controlled devices in terms of their number and types of registered accounts, the number of apps reviewed, and the intervals between the installation times of apps and their review times. This suggests that the constraints associated with ASO work provide opportunities to detect even organic workers.

%\newmaterial{We model each device as a discrete Markov chain built from app transitions, and found that workers have a longer mixing time, and shorter hitting times with respect to the play store in such chains.}

To validate this, we leverage our findings and the RacketStore-collected data to develop features that model the interaction of a user with a device and the engagement of the user with the apps installed on the device. We found that supervised learning algorithms trained with these features distinguish between apps installed for promotion purposes and those installed for personal use, thus detect incentivized installs and fake reviews, with an F1-measure of 99.72\% and AUC over 0.99. Further, our classifiers detect worker-controlled devices with an F1-measure of 95.29\% and AUC of 0.95.

We found that 69.1\% of the worker devices that we analyzed have organic-indicative behaviors, while the remaining devices were seemingly used exclusively for app promotion activities. This suggests that our device and app-engagement features can train classifiers to accurately detect not only promotion-dedicated devices but even elusive organic ASO efforts that hide low levels of app promotion activities among regular, personal use of devices and of the apps installed therein.

%Workers who attempt to emulate the genuine interest of regular users to evade detection by the proposed classifiers will need to keep each promoted app installed longer, interact more with the app, and significantly reduce the number of posted reviews. However, the higher work load, reduced profits and increased exposure to malware reduce the appeal of this strategy ($\S$~\ref{sec:discussion}).

We note that to protect user privacy, the proposed classifiers can execute directly on the user device (e.g., implemented into the Play Store app). Locally-running classifiers can access sensitive app and device usage data and do not need to report it remotely ($\S$~\ref{sec:discussion}).

In summary, we introduce the following contributions:

\begin{compactitem}

\item
{\bf RacketStore}.
We develop a platform to collect information about the interaction of users with their Android devices and the apps installed therein, with user consent. RacketStore was compatible with 298 device models from 28 Android manufacturers [$\S$~\ref{sec:racketstore}]. \newmaterial{The RacketStore code is available at~\cite{RacketStore}}.

\item
{\bf App and Device Use Measurements}.
We present measurements from a study of the device and app use of regular users and ground truth ASO workers, through a deployment of RacketStore on \FPeval{\result}{clip(\numregular+\numworkers+\numexchangers)} \numprint{\result} unique devices [$\S$~\ref{sec:collection}]. We build datasets of app and device usage, integrated with Google Play reviews and VirusTotal analysis. We present findings from this data in the context of feedback obtained from participants during a follow-up discussion [$\S$~\ref{sec:findings}].

\item
{\bf Fraud Detector and Classifier}.
We introduce novel features that model the user interaction with devices and installed apps and use them to train classifiers to detect ASO activities [$\S$~\ref{sec:app}] and worker-controlled devices [$\S$~\ref{sec:device}]. We report differences in app and device-engagement for workers and regular users that explain the accuracy of the classifiers.

\end{compactitem}

\section{System Model}
\label{sec:model}

\begin{figure}
\centering
\includegraphics[width=0.97\columnwidth]{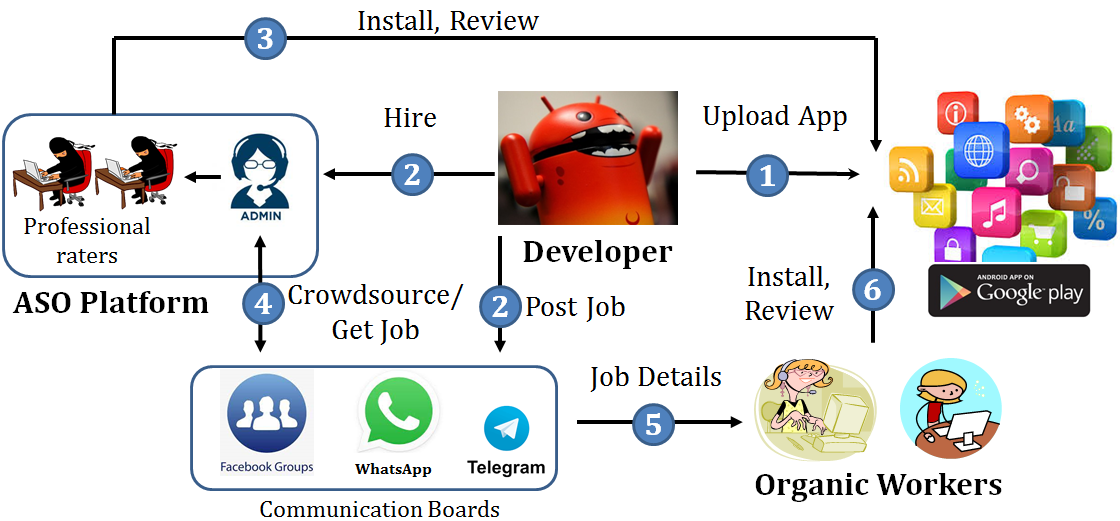}
\caption{System model. Developers recruit ASO platforms to promote their apps in app stores. ASO platforms leverage in-house, dedicated workers, and organic workers accessed through communication boards to install and review apps.}
\label{fig:model}
\vspace{-15pt}
\end{figure}

We consider the ecosystem depicted in Figure~\ref{fig:model}. In the following we describe its main components.

\noindent
{\bf The App Store and Consumers}.
We focus our work on the Google Play app store~\cite{googleplay}. Consumers use the pre-installed Play Store app to search and install other apps on their Android devices. A consumer can register multiple accounts on an Android device, including Gmail and other services. The consumer is then able to post reviews for an app, from all the Gmail accounts registered on the device where the app was installed.

\noindent
{\bf App Developers}.
Developers upload their apps on the Play Store~\cite{googleplay}. To monetize these apps while facing intense competition, they need to achieve top-5 rank in keyword searches~\cite{WhiteHatASO}. Some of the factors with most impact on search rank are the number of installs and reviews, and the aggregate rating of the app: 80\% of consumers check reviews and ratings before installing an app ~\cite{CPVW12}, and a 1-star increase in aggregate rating was shown to increase app store conversion by up to 280\%~\cite{Apptentive}.

\noindent
{\bf ASO Organizations}.
Many developers hire specialized app search optimization (ASO) organizations to improve the search rank of their apps. While some ASO organizations are white hat~\cite{WhiteHatASO}, providing advice on e.g., app naming and keyword optimization, others provide illegal, banned or discouraged services that include installing an app on many devices and keeping it installed for prolonged intervals, a.k.a., {\it retention installs}, and writing fake reviews with high ratings.

We consider the ASO organizations studied in~\cite{RHRAC19}, see Figure~\ref{fig:model}, that employ combinations of (1) ASO admins who organize and coordinate communities of workers, and act as intermediaries between developers and workers, (2) professional workers who use multiple devices and accounts dedicated to ASO work, and (3) organic workers who blend product promotion with personal activities from their devices and accounts. In the following we informally use the terms organic device and account, to denote devices and accounts used by organic workers.

Developers can either directly hire ASO organizations or post jobs in online boards dedicated to ASO work, see next.

\noindent
{\bf ASO Communication Boards}.
Communications between developers, admins and crowdsourced workers often occur through dedicated online boards in e.g., Facebook, WhatsApp, Telegram~\cite{RHRAC19}, see Figure~\ref{fig:model}. In this paper we recruited participants through Facebook groups that we identified using Facebook's search functionality (using keywords that include reviews, google reviews, app reviews, app installs, android promotion) to identify relevant groups.

We found 11 public and 5 closed groups that matched our criteria. We became members of the public groups and sent requests to closed groups which were all accepted. These groups had 86,718 members (Min = 354, Max = 26896, M = 2840.5, SD = 6787.96) in total.
%
%Figure~\ref{fig:fbgroup:members} (Appendix~\ref{appendix:facebook:fraud}) shows the number of members in each group.
%
We detail our recruitment process in $\S$~\ref{sec:collection}.

\section{Measurements Infrastructure}
\label{sec:racketstore}

\begin{figure*}
\centering
\includegraphics[width=0.75\textwidth]{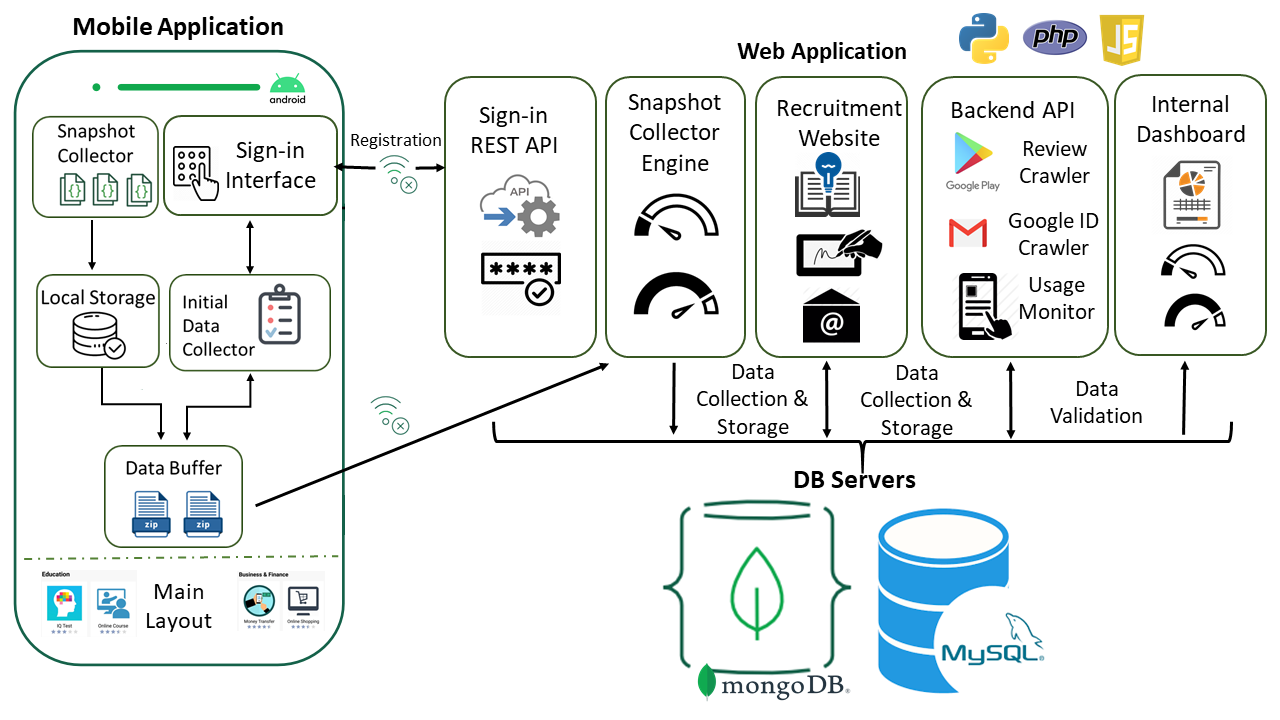}
\vspace{-10pt}
\caption{RacketStore architecture consists of a mobile app installed by participants and a back-end server that collects and aggregates snapshots reported by deployed apps.}
\label{fig:system}
\vspace{-5pt}
\end{figure*}

We have built the RacketStore platform to measure and compare the app and device usage of ASO workers and regular users. RacketStore consists of a mobile app to be installed by study participants on their Android devices, a web app to collect and validate data from the installed app, and database servers to store the data, see Figure ~\ref{fig:system}. In the following we describe the main components of RacketStore.

\noindent
{\bf RacketStore Mobile App}.
We have developed the RacketStore app in Android to help us investigate fraudulent and honest behaviors of Google services users. The app needs to be installed by study participants on their devices. Upon first start-up, the app displays the consent form (see Appendix~\ref{appendix:consent} for excerpts) which the participant needs to approve. Then, to comply with the Google anti-abuse policy~\cite{antiabuse}, the app asks for explicit consent of our privacy policy (Figure~\ref{fig:screenshot1} in Appendix~\ref{appendix:consent}) then shows an in-app disclosure of the data being collected (Figure~\ref{fig:screenshot2}). In the following we detail the main components of the rest of the app.

The {\bf sign-in interface} asks the participant to enter a unique {\it participant ID}, a 6-digit code, that we send upon recruitment ($\S$~\ref{sec:collection}) through a \newmaterial{different channel}, i.e., e-mail or Facebook messenger. This code serves the dual goal of preventing RacketStore use by non-recruited users, and of allowing us to match data and send payments to the correct participants. The passcode is given only after the user has accepted to participate in our study and has agreed to the data collection process. RacketStore does not collect any information if the user has not entered the 6-digit passcode. Upon sign-in, the app generates the {\it install ID}, a 10-digit random identifier.

The {\bf initial data collector} module operates once the app has been installed \newmaterial{and the user has used the sign-in interface to enter the participant ID.} It retrieves the list of other apps installed on the device, and device information including Android API version, device model, manufacturer, and Android ID~\cite{androidid}.

Following the installation of the RacketStore app, the {\bf snapshot collector} module periodically collects information with two levels of granularity, slow and fast. The slow snapshot collector is triggered by an alarm every 2 minutes, and collects (1) {\it identifiers}: Install ID, participant ID, and Android ID, (2) {\it registered accounts}, the accounts registered on the device across different services, (3) {\it device status},  i.e., save mode status (on/off), and (4) {\it stopped apps}, the list of stopped apps. Starting with Android 3.1 all applications upon installation are placed in a stopped state: the application will only run after a manual launch of an activity, or an explicit intent that addresses an activity, service or broadcast. The user can also manually force stop an app.

The fast snapshot collector module further activates every 5s and collects (1) {\it identifiers}, i.e., install ID and participant ID, (2) the {\it foreground app} currently running on the device foreground, (3) the {\it device status}, i.e., the screen status (on/off) and battery level, and (4) {\it app install/uninstall events}, i.e., deltas between the current and previously reported sets of installed apps. For each installed app we collect the install time, the last update time, the required permissions and the MD5 hash of the app apk file.
%
%In Appendix~\ref{appendix:findings:utilization} we show that at this collection rate RacketStore had minimal impact on device CPU and battery utilization.

RacketStore requires participants to explicitly grant two permissions, \texttt{PACKAGE\_USAGE\_STATS} and \texttt{GET\_ACCOUNTS} ~\cite{androidpermission}. \newmaterial{Participants can accept any subset of the requested permissions. If they do not grant a permission, we do not collect the corresponding data.} \newmaterial{RacketStore also uses install-time permissions (\texttt{GET\_TASKS}, \texttt{RECEIVE\_BOOT\_COMPLETED}, \texttt{INTERNET}, \texttt{ACCESS\_NETWORK\_STATE}, and \texttt{WAKE\_LOCK}) that are automatically granted when the app is installed.} The RacketStore app was approved as compliant by the Play Store.

\noindent
{\bf Data Buffer Module: Snapshot Processor}.
The \textit{data buffer} module (see Figure~\ref{fig:system}) leverages the device storage to process both types of snapshots. The snapshot data is written to different accumulating files depending on the snapshot type. When the slow snapshot accumulation file reaches 8KB and the fast snapshot file reaches 100KB, this module compresses the file and creates a new accumulation file to store the following snapshots. We selected these threshold values based on the observed battery and bandwidth consumption, which we sought to minimize.

The slow snapshot alarm that fires every 2 minutes looks for any existing compressed files in the mobile app's directory and sends them to our server. To enable resilient communications, upon file reception, the server returns the crypto hash of the received data in order for the mobile app to validate the transfer with its own hash calculation. If the hashes are equal, the data buffer module deletes the file.

\noindent
\newmaterial{
{\bf Device Compatibility}.
The RacketStore app was targeted for devices with Android version 9 (compile SDK and target SDK version are Android Pie, API level 28) and is compatible with devices with Android version at or above Lollipop (min SDK version = 5, API level 21). In our deployment study ($\S$~\ref{sec:collection}), RacketStore was compatible with 298 unique device models from 28 Android device manufacturers. The top 5 most popular Android manufacturers were Samsung, Huawei, Oppo, Xiaomi, Vivo).}

\noindent
{\bf RacketStore Web App}.
We have built a web app that supports RacketStore on the server side, see Figure~\ref{fig:system}. The Sign-in component processes registration requests from the client app, interacts with the Mongo database where the credentials are stored and sends the response back to the client. The snapshot collector engine receives the compressed snapshot files from the app, decompresses, and inserts them into the database. The backend component consists of two subsystems: (1) a review crawler that scraps reviews from Google Play given an app name and a device model and (2) a Google ID crawler that maps Gmail accounts to a unique Google ID (see $\S$~\ref{sec:data}). The internal dashboard allows researchers to monitor the data collection process, and test and validate the data sent from the app to the server. Our stack is Linux-based and 
built on Python, PHP, JavaScript, MongoDB, and MySQL. The recruitment website is an informative website where we offer information to participants about our study, ask for consent, and collect their emails to send instructions on how to proceed ($\S$~\ref{sec:collection}).

\noindent
\newmaterial{
{\bf Security and Privacy Risks of RacketStore}.
RacketStore uses TLS to encrypt all collected data while in-transit. We also securely store all participant data on a server that is accessible from only 4 IP addresses in our campus. Further, RacketStore does not collect any information if the user has not entered a 6-digit passcode. Each study participant received a unique, random code. Participants can deny access to any of the two permissions requested by RacketStore.}

\newmaterial{
RacketStore minimizes collected Personal Identifiable Information (PII) by only collecting PII that is needed for the study. Table~\ref{tab:pii} summarizes the participant PII that RacketStore collected, the reasons for the collection and how long it was stored.}

\section{Data Collection}
\label{sec:collection}

In order to collect app and device usage, we deployed RacketStore with a group composed of both ASO workers and regular Google Play users. In this section we detail our recruitment process and discuss ethical considerations.

\noindent
{\bf Recruitment of ASO Workers}.
We recruited ASO workers from Facebook groups that we found to be dedicated to product promotion (see $\S$~\ref{sec:model}). Specifically, we posted calls to recruit participants who would be able to install and provide reviews. Many group members commented on our posts, expressed their interest, and asked us to communicate with them over the Facebook inbox for further details. We contacted such members over the Facebook inbox. We shared with them the details of the recruitment instructions which we include in Appendix~\ref{sec:collection:materials}. We asked them to reply if they were interested and also to answer screening questions, i.e., confirm that they have posted paid reviews, specify how many devices and accounts they have and on how many devices they can install RacketStore.

We sent to each prospective participant, their participant ID ($\S$~\ref{sec:racketstore}) and a YouTube video that explains how to sign up for RacketStore. We have received 672 installs from 549 unique worker-controlled devices. 

\noindent
{\bf Recruitment of Regular Users}.
We have recruited regular Android device users through commercial advertisements on Instagram (see Figure~\ref{fig:ad} in Appendix~\ref{sec:collection:materials}) that point to a landing page that explains the study (see Figure~\ref{fig:landing} Appendix~\ref{sec:collection:materials}). We chose Instagram in order to \newmaterial{minimize the exposure of Facebook groups of ASO workers to our ads}.

We posted the ads intermittently between December 17, 2019 and April 15, 2020, spending a total of \$79.23. Since cultural differences could affect patterns in mobile device use~\cite{WK05,culturalmobile}, we targeted regular users of similar demographic with the recruited ASO workers. Concretely, we used Facebook's audience creation functionality to ensure that our ads were shown only to mobile devices of Instagram users who are from the countries of the above workers, are between 18 and 40 years old, speak English, and show interests related to Google Play and Android applications as specified on their Facebook profiles.

According to the Facebook Ads Manager, our ad was shown a total of 136,022 times and reached 61,748 users. 2,471 of these users clicked on the Instagram ad and made it to our landing page. The landing page introduces our study, explains the payment method (i.e., Paypal, Bitcoin or Litecoin), presents the consent form, and allows visitors to either withdraw or sign up. To sign up, a visitor needs to acknowledge and agree with the terms and conditions that are included in the consent form explaining in detail the information that we would collect from their phones (see $\S$~\ref{sec:methods:ethical} for more details). If the visitor consents, the landing page asks them to register in the study by submitting their email address.

\newmaterial{
Of the consenting visitors, we have filtered out those who claimed to have written paid reviews in Google Play (question 1 in the recruitment message, see Appendix~\ref{sec:collection:materials}) and who claimed to be administrators (question 6).}

To each of the remaining 614 visitors, we sent an automatic confirmation email along with the Google Play link to download the RacketStore app ($\S$~\ref{sec:racketstore}) and a six-digit unique participant ID that the participant would need to type-in to the app. RacketStore received 233 installs from these participants.

\noindent
{\bf Participant Payments}.
We paid each participant who installed RacketStore on a per device basis: \$1 to install the app and \$0.2 for each day on which the participant kept the app installed. The process of registering in the study, providing consent and installing RacketStore takes an average of 3 minutes. Participation does not require any subsequent user interaction. We paid participants in our follow-up study \$5 for each 15 minutes of their time.

Overall, we recruited a ground truth set of 587 ASO workers and 233 regular users. The participants used IP addresses from Pakistan (420), India (210), Bangladesh (148), USA (10) and other countries from Africa, Asia, South America and Europe (15). The distribution of Worker (W) and Regular (R) participants for the most represented countries was as follows: Pakistan (W: 364, R: 56), India (W: 57, R: 153), Bangladesh (W: 143, R: 5) and USA (W: 8, R: 2). \newmaterial{We note that IP addresses can only provide an approximate measure of geolocation~\cite{MV06}. The RacketStore app did not collect location information from participant devices thus we cannot corroborate this information.}

\subsection{Ethical Considerations}
\label{sec:methods:ethical}

\newmaterial{
Some ASO work is considered unethical according to several ethical frameworks, and many ASO workers belong to low-paid vulnerable groups. This is why our study took utmost care to follow the best ethical practices for conducting sensitive research with vulnerable populations~\cite{BBMR17}.} We did not use any deception in our study. Participation in this study was completely voluntary. We did not ask any participant to write reviews for any app, including for the RacketStore app. We included the consent form both in the landing page for our study and in the RacketStore app (see excerpts in Appendix~\ref{appendix:consent}). The consent form explicitly mentions the identity of the researchers, the research objectives, the data that we wanted to collect, and the potential impacts on participants, including risks. Our team members were also available to explain this to the participants if needed. Each participant needed to explicitly provide consent. \newmaterial{The full study procedure was examined and approved by the Institutional Review Board of our university (IRB-19-0392@FIU)}.

In Appendix~\ref{appendix:ethics} we further discuss the privacy policy and permissions requested by the RacketStore app, our data protection procedure, and participant compensation.

\section{Data}
\label{sec:data}

We now detail the data collected by RacketStore from 943 devices between October 2019 and April 2020.

\noindent
{\bf Snapshot Fingerprinting and Coalescing}.
We used a combination of the install ID, participant ID and Android ID collected by RacketStore, along with its install interval to address repeat installs and suspected incompatibilities of the RacketStore app, see Appendix~\ref{appendix:fingerprinting}. After this process, we identified \FPeval{\result}{clip(\numregular+\numworkers+\numexchangers)} \numprint{\result} unique devices: \FPeval{\aresult}{clip(\numworkers+\numexchangers)} \numprint{\aresult} devices controlled by ASO workers and \numregular~ devices controlled by regular participants recruited through Instagram ads ($\S$~\ref{sec:collection}).

We have then collected a total of %592,045 
\numprint{\numslowsnaps} slow snapshots and %57,770,204 
\numprint{\numfastsnaps} fast snapshots ($\S$~\ref{sec:racketstore}).

\noindent
{\bf Google Play Review Dataset}.
The review crawler of the Backend component of  RacketStore's web app (see Figure~\ref{fig:system}) collects reviews posted for apps installed on participant devices every 12 hours. For each of these apps, we collected the most recent reviews by querying Google Play for reviews sorted by timestamp. The first time an app was processed, we collected reviews until hitting a threshold of 100,000 reviews. In subsequent collection efforts, we collected the most recent reviews until finding a previously collected review. This procedure allowed us to collect reviews ``live'' as soon as our RacketStore mobile application discovered a new app on a participant's device.

\newmaterial{Further, we collected reviews posted by accounts registered on participant devices. This process uses the Google ID crawler component of the Backend API in RacketStore’s web app (Figure~\ref{fig:system}), maps e-mail accounts to Google IDs. To achieve this, for each Gmail account registered on a device, the ID crawler issues requests to Gmail’s email search functionality. This is because we found that responses of Gmail’s email search functionality embed the Google ID. We then list all the apps installed on each device, that are hosted on the Play Store. Then, for each such app, we search the Google IDs corresponding to accounts registered on the device, among the Google IDs collected by the above review crawler from the Play Store for the app.}

We collected \numprint{\numreviews} reviews from \numprint{\totalapps} apps installed on participant devices. Each review includes metadata, e.g., the user's Google ID, posting timestamp (1s granularity) and rating.

We also collected 217,041 reviews posted by the 10,310 Gmail addresses registered on the worker-controlled devices of the participants in our studies, with participant consent ($\S$~\ref{sec:methods:ethical}). We have reported to the Google Vulnerability Reward Program (VRP, \newmaterial{issue ID: 156369357}) the functionality that we used to collect this data. \newmaterial{More specifically, our finding that responses of Gmail's email search functionality embed the user's Google ID. This allows a third party with a Gmail address, to obtain the account's Google ID, then determine if the account has posted a review in the Play Store for any app of interest.} \newmaterial{Google responded that this is ``intended behavior'' and made the decision ``not to track it as a security bug''.} We notified study participants about this data collection ($\S$~\ref{sec:methods:ethical}).

\section{Device Usage Measurements}
\label{sec:findings}

We use the data collected from the participant devices ($\S$~\ref{sec:data}) to investigate differences between workers and regular users in terms of the accounts registered and the apps installed on their devices.

\newmaterial{We used the Kolmogorov–Smirnov (KS) test to compare the distributions of the features, and non-parametric and parametric ANOVA (Analysis of Variance) to check for significant differences in the means across groups. The reason for this is that after performing the Shapiro test, we can not claim normality for any of the features ($p$-value$<0.05$). Similarly, after performing the Fligner-Killen test, we found significant differences in the variances for all the features ($p$-value$<0.05$). Thus, we have also performed non-parametric ANOVA since normality is not assumed. We report the result of the three approaches.}

%Appendix~\ref{appendix:features} includes further analysis of device-use features, e.g., daily used apps and app permissions.

\subsection{Participant Engagement}

\begin{figure}
\vspace{10pt}
\includegraphics[width=0.95\columnwidth]{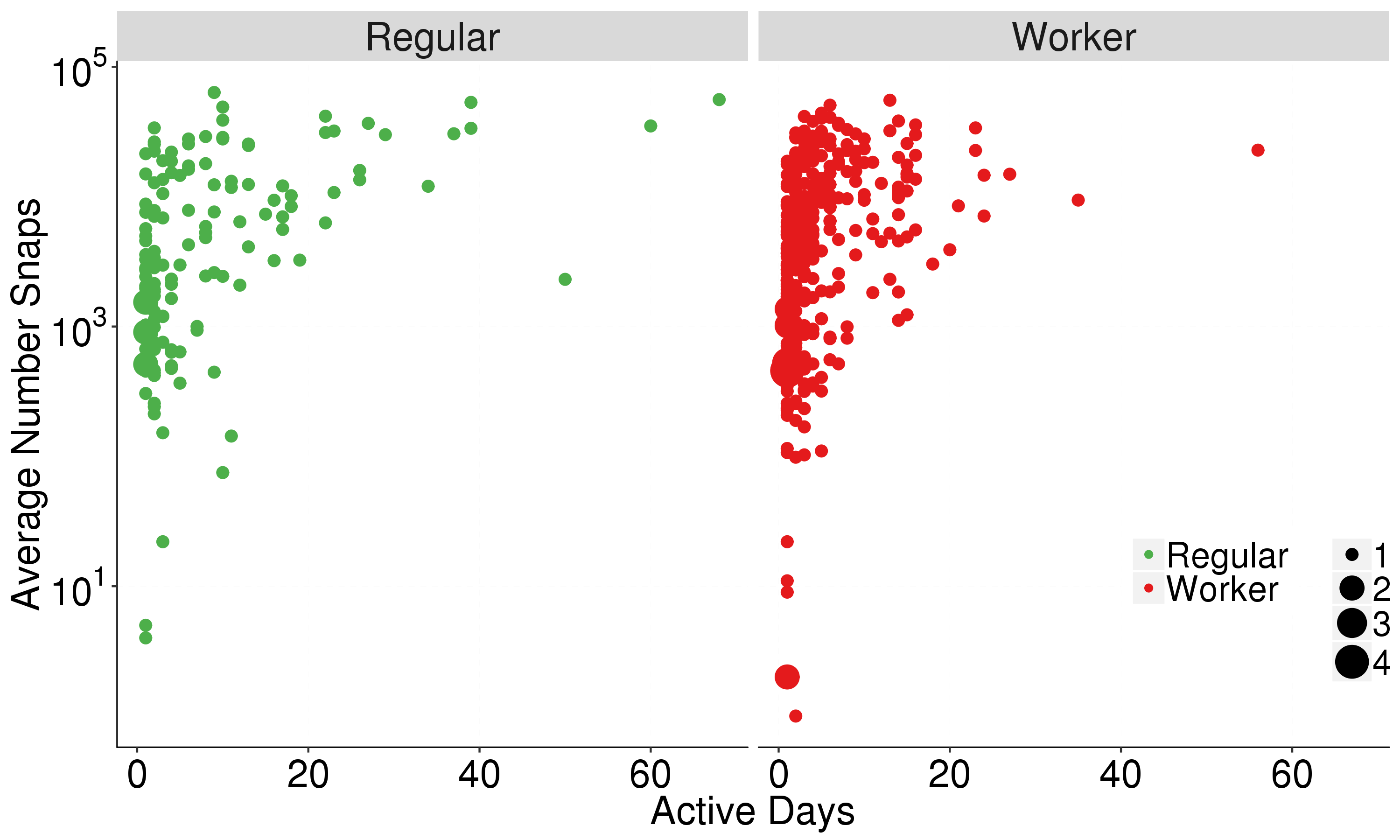}
\vspace{-10pt}
\caption{Scatterplot of average number of snapshots collected per day vs active days over regular (green), and worker (red) devices. Dot size indicates the number of overlapping devices. Most devices report at least 100 snapshots per day.}
\label{fig:scatterAverage}
\vspace{-5pt}
\end{figure}

Figure ~\ref{fig:scatterAverage} shows the scatterplot of the average number of snapshots per day vs. the number of active days over the participant devices. Larger dots denote multiple overlapping devices. The average number of daily snapshots collected from regular devices is 9,430.71 (M = 3,097.67, SD = 12,789.14, max = 63,452) and from worker devices is 8,208.10 (M = 3,669, SD = 10,303.42, max = 55,281.38). The maximum number of snapshots per day is 55,281.38. We observe that 529 devices have reported at least 100 snapshots per day.

\subsection{Registered Accounts}
\label{sec:findings:accounts}

\begin{figure}
\includegraphics[width=0.89\columnwidth]{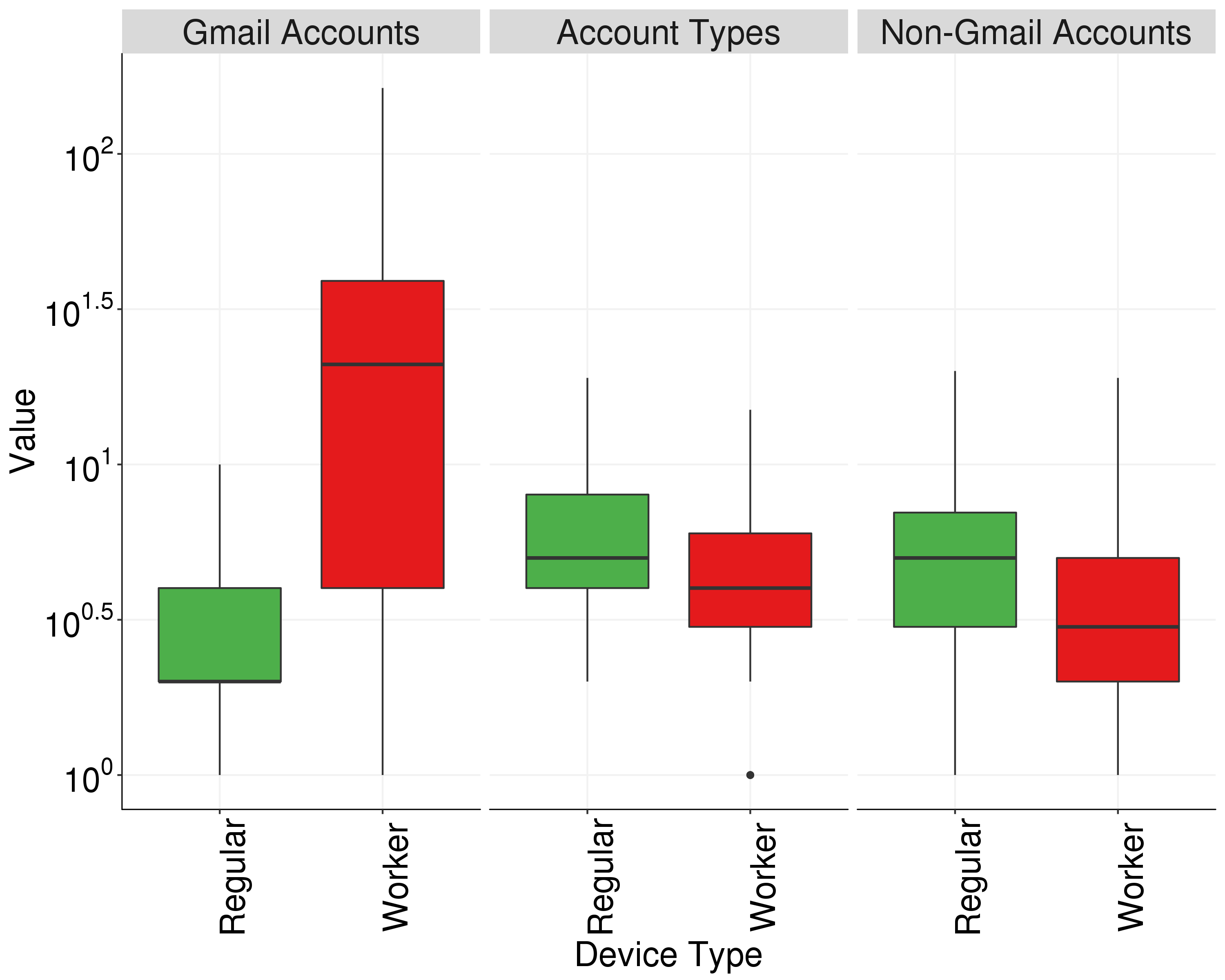}
\vspace{-10pt}
\caption{Comparison of the number and types of accounts registered on devices controlled by ASO workers and regular participants. Worker devices tend to have more Gmail accounts, but fewer account types and non-Gmail accounts than regular devices.}
\label{fig:accounts}
\vspace{-15pt}
\end{figure}

To post a review, a user needs to have a Gmail account. For one app, a single review can be posted from any Gmail account. We now investigate differences in the number and types of accounts controlled by workers and regular users. We expect that workers will have more Gmail accounts registered on their devices than regular users since this impacts the number of reviews that they can post.

Figure~\ref{fig:accounts} (left) shows the number of Gmail accounts registered on the 145 regular and 390 worker devices that have reported such information. The other participants either did not grant this permission or the server did not receive enough snapshots from their devices. We found significant differences between regular user and worker-controlled devices. Worker devices have an average of 28.87 accounts registered per device (M = 21, SD = 29.37); 13 worker devices have more than 100 Gmail accounts registered, with a maximum of 163 accounts per device. In contrast, regular devices have a maximum of 10 accounts registered (M = 2 , SD = 1.66). \newmaterial{The KS test, and parametric and non-parametric ANOVA found statistically significant differences between workers and regular users ($p$-value$<0.05$).}

Figure~\ref{fig:accounts} (center) shows the number of different account types registered on participant devices. On average, regular devices have registered accounts for 6 services (max = 19), mostly for different social networks (Facebook, WhatsApp, Telegram, etc). In contrast, worker devices have accounts mainly for Google services and other services useful for ASO work, e.g., dualspace.daemon (to enable installation of the same app multiple times) and freelancer (to find work). \newmaterial{Both KS and ANOVA analyses reveal significant differences between workers and regular users ($p-$value$<0.05$) in terms of their numbers of non-Gmail accounts.}

We have followed up with several participant workers. Six of the ASO workers who replied claimed that they personally own only 1-4 Google accounts. For instance, one worker said ``{\it I have two accounts.  One account is mine, another is my mom's.}'' Four other workers however claimed (and we verified) to control between 10 to 50 Gmail accounts, and one claimed to have ``{\it many accounts}''.

\noindent
{\bf Summary of Findings}.
We confirmed that participant ASO workers have registered significantly more accounts on their devices than regular users. Workers have however less diversity in the online sites for which they registered accounts. Their accounts are specialized for ASO work, focusing on Gmail and Dualspace that enable them to install and review a single app multiple times.

\subsection{Installed Apps}
\label{sec:findings:installed}

\begin{figure}
\includegraphics[width=0.89\columnwidth]{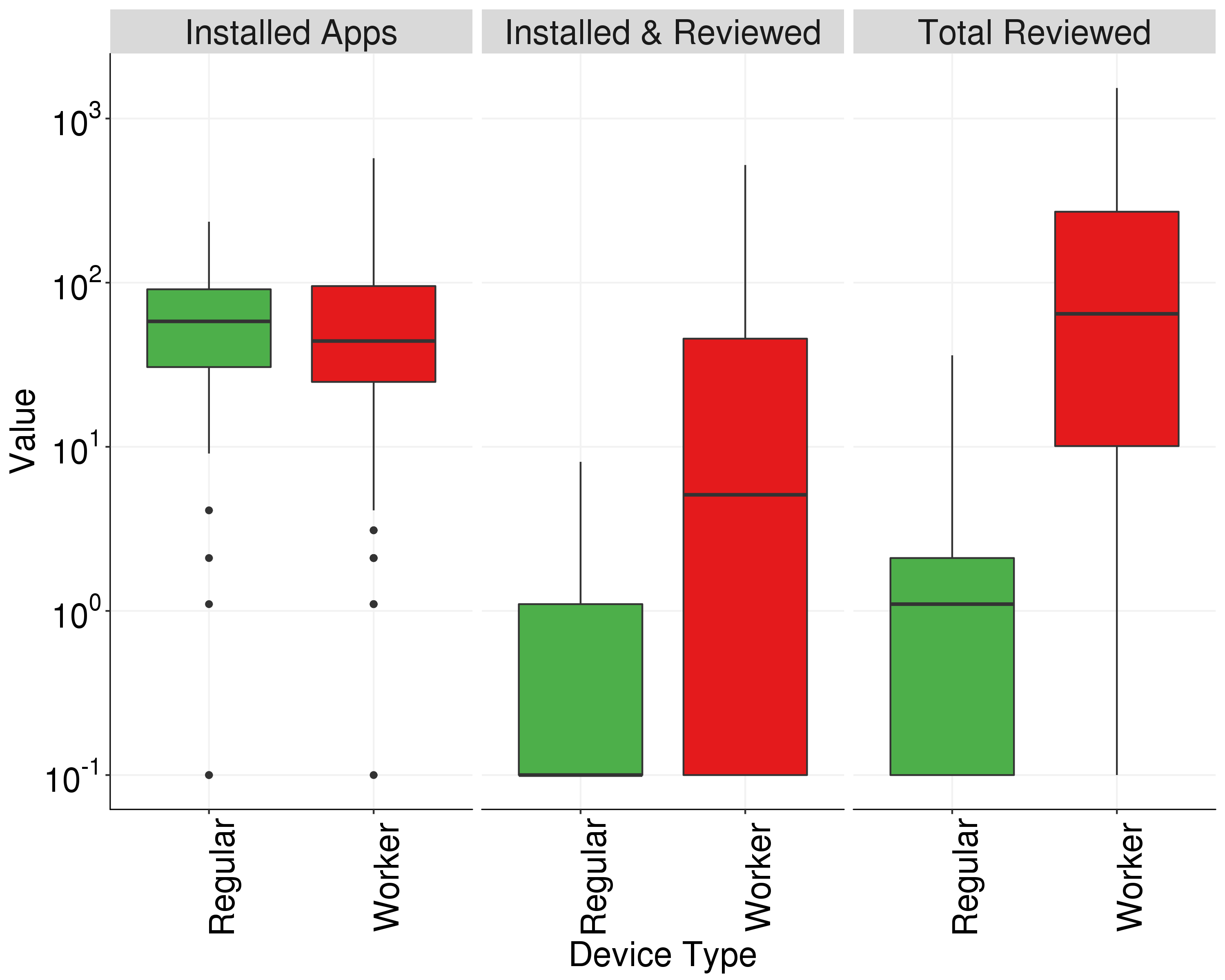}
\vspace{-10pt}
\caption{Number of installed apps (left), installed and reviewed (center), and total number of reviews posted from all accounts registered (right). We see dramatic differences between worker and regular devices in the apps reviewed from all their registered accounts.}
\label{fig:install:review}
\vspace{-15pt}
\end{figure}

We now investigate the hypothesis (inspired from~\cite{FFLMSV20}) that workers and regular users differ in the manner in which they interact with installed apps.

\noindent
{\bf Apps Installed and Reviewed}.
Figure~\ref{fig:install:review} compares the distribution of the number of installed apps (left), the number of apps installed and reviewed (center), and the total number of apps reviewed from any account registered (right) for the 143 regular and 400 worker devices that reported this data. We observe dramatic differences between worker and regular devices in terms of the total number of reviews posted from registered accounts: On average, a worker device is responsible for a total of 208.91 reviews, while a regular user device has only posted an average of 1.91 reviews. We found 11 worker-controlled devices each responsible for more than 1,000 total reviews. In contrast, the maximum number of total reviews from a regular device is only 36. \newmaterial{We found statistically significant differences between workers and regular users via KS and parametric and non-parametric ANOVA ($p$-value$<0.05$).}

We observe an average of 65.45 and 77.56 apps installed on regular and worker devices respectively. \newmaterial{KS reported significant differences in the distributions ($p$-value$= 0.008$), but ANOVA did not find a statistically significant difference ($p$-value$=0.301$)}. This is expected since the number of installations is limited by the device resources. However, on average, worker devices have posted reviews for 40.51 of the currently installed apps, while regular devices did it for an average of 0.7 apps.

\begin{figure}
\vspace{10pt}
\includegraphics[width=0.89\columnwidth]{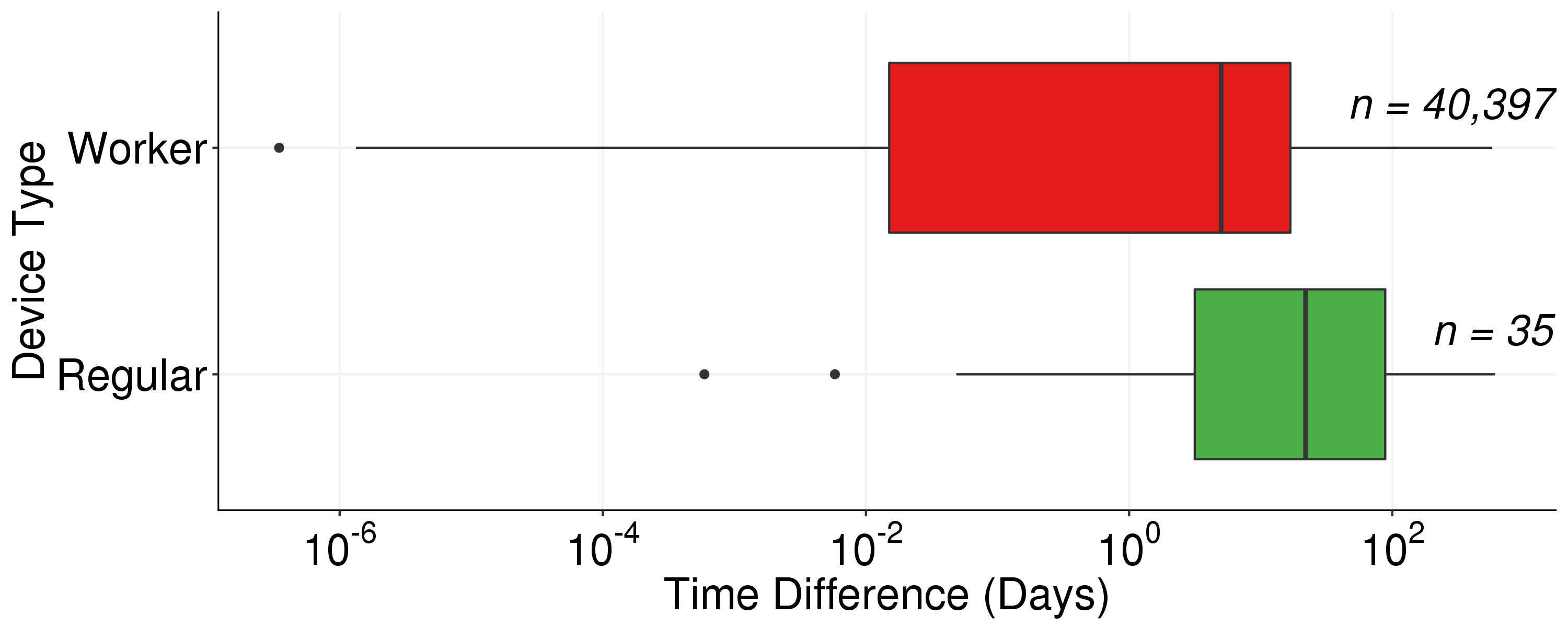}
\vspace{-10pt}
\caption{Distribution of time between app install and app review, for regular and worker devices. Each point is one review. Unlike regular users, worker-controlled accounts post many more reviews and tend to do it soon after installation. 13,376 of the reviews from worker accounts were posted after at most one day after installation.}
\label{fig:install:review:time}
\vspace{-15pt}
\end{figure}

\noindent
{\bf Install-to-Review Time}.
\newmaterial{The Android API reports the time of each app's last install. There are thus two cases. First, the time when an app was installed on a participant device is before the time when the participant reviewed the app. This is likely the case where the collected review is from the currently installed version of the app on that device. Second, the install time is after the review time. This implies that the review we collected is from a previous install of the app. For our analysis, we only consider the first case.}

Figure~\ref{fig:install:review:time} shows the distribution of the time between app install and app review, for regular and worker devices. Each point is one review. An app can be reviewed from multiple accounts registered on the same device; each such review is a different point. To compute the install-to-review time, we used the Android API to get the installation time, and our review crawler to get the review timestamp. However the Android API only retains the last installation time of an app in a device. We have not considered reviews whose install-to-review times were negative, since they are the result of a past install.

We observe substantial difference in the number of reviews posted from accounts registered on worker vs. regular devices for apps that provided an installation time: accounts on regular devices only wrote 35 reviews, while those on worker devices posted 40,397 reviews. \newmaterial{Both ANOVA and KS tests found statistically significant differences between the two groups ($p-$value<0.05).}

Further, workers tend to review apps much sooner after installation. 13,376 of the 40,397 reviews posted from the accounts registered on worker devices were posted after at most one day after the app was installed. Workers register an average of 10.4 days of waiting time between installation and review (M = 5.00 days, SD = 13.72 days, max = 574 days). We have observed 25 cases with waiting times longer than 100 days and 4 cases of reviews posted after more than 1,000 days from 2 workers and for 2 apps (Facebook and Easypaisa). These rare cases of prolonged waiting times are expected of apps used for personal purposes.

In contrast, only 4 out of the 35 reviews posted from the accounts of regular users were posted after at most one day after install. Regular users also wait for 85.09 days to post a review on average (M=21.92 days, SD=140.56 days, max=606.11 days) with only 12 users waiting less than 12 days to post a review. This longer waiting time is consistent with a review activity that proceeds from a previous interaction to form a judgment, and is inconsistent with paid promotion services. \newmaterial{KS and ANOVA found statistically significant differences ($p-$value<0.05).}

We have observed 25 cases for worker devices having waiting times longer than 100 days, and 4 cases of reviews posted after more than 1,000 days (from 2 workers for Facebook and 2 for Easypaisa). These rare cases of prolonged waiting times may be indicative of apps used for personal purposes.

\begin{figure}
\vspace{10pt}
\includegraphics[width=0.89\columnwidth]{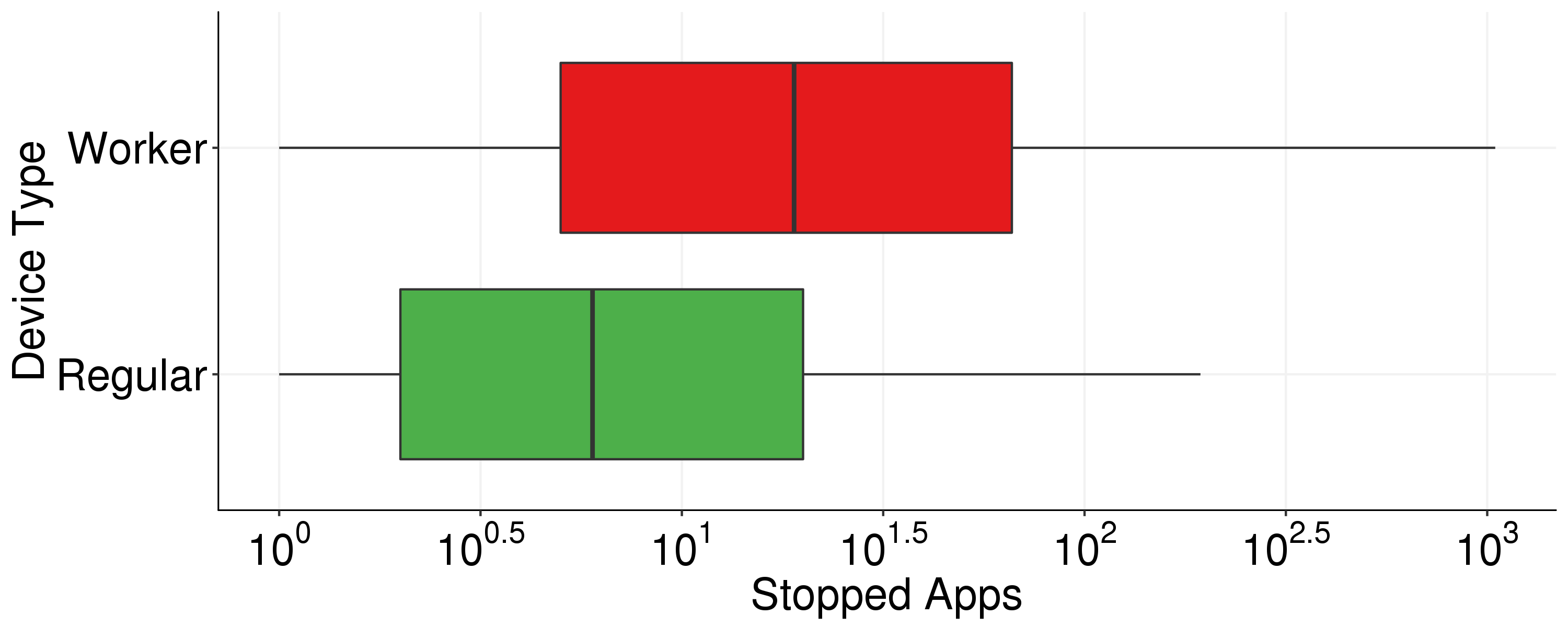}
\vspace{-10pt}
\caption{Boxplot of stopped apps for regular and worker devices. Worker devices tend to have more stopped apps, but we also observe substantial overlap with regular devices.}
\label{fig:stopped}
\vspace{-15pt}
\end{figure}

\noindent
{\bf Stopped Apps}.
We have further studied the number of apps that are stopped on the devices of regular and worker participants. A freshly installed app is in a stopped state until the user opens it for the first time. Android devices also allow users to stop apps, instead of uninstalling them. Figure~\ref{fig:stopped} shows that some worker devices have significantly more stopped apps than regular devices. \newmaterial{KS test and ANOVA found statistically significant differences between workers and regular users ($p$-value$<0.05$)}. We conjecture that this occurs because ASO workers (1) often do not open the apps that they install in order to promote, and (2) even if they open them and need to keep them installed, e.g., to provide {\it retention installs} ($\S$~\ref{sec:model}), they prefer to stop apps that misbehave.

We followed up with several participants to clarify this point, i.e., whether they stop apps and why. Eight workers claimed to never stop apps, which suggests reason \#1 applies. One worker however admitted reason \#2 applies, i.e., ``{\it The quality of some apps was bad, I stopped those apps}''. Another claimed that limited storage is to blame, ``{\it Sometimes the apps get hanged due to a lack of storage}''.

\noindent
{\bf Third-Party App Stores}.
We observed that some participant devices had apps installed that were not available in Google Play. We followed up with participants to ask if they install apps from other app stores. The workers conjectured that Google does not host such apps because they violate Google's policy, e.g., ``{\it Google's policy prohibits the use of such apps}, or because they are not secure.

One worker admitted to have installed apps from third-party stores, i.e., ``{\it The client gives us a link, we go and install that app}''. Three other workers claimed to install apps from other app stores for personal reasons, e.g., to play games (Dream11) or avoid subscription fees (Netflix, Hotstar) by installing {\it modded} apps~\footnote{A {\it mod app} is a modified version of an original apk, not signed by the original developers. A modded app may have additional features, unlocked features, and unlimited in-app currency.}:

``{\it I use a modded version of the apps that are not in the Google Play Store. You do not have to open an account to use these. For instance, Netflix or Hotstar apps charge a subscription fee every month. But I don't have that much money so I install the modded version. By doing this I get premium access for free. }''

\begin{figure}
\vspace{10pt}
\includegraphics[width=0.75\columnwidth]{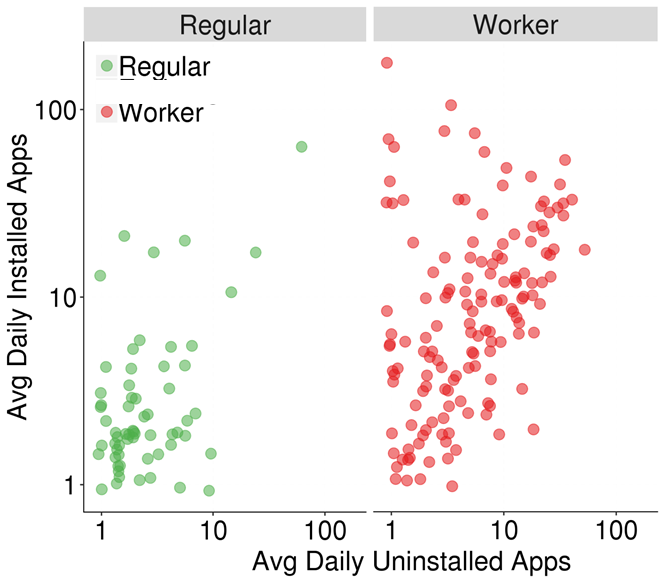}
\vspace{-10pt}
\caption{App churn: Scatterplot of average number of daily installs vs. average number of daily uninstalls (log scale) for regular and worker-controlled devices. Each dot is one device. The app churn of most regular devices is less than 10 apps per day, while for many worker devices it is above 10 apps per day.}
\label{fig:install:uninstall}
\vspace{-15pt}
\end{figure}

\noindent
{\bf App Churn: Install and Uninstall Events}.
Figure~\ref{fig:install:uninstall} shows the average number of daily install events and daily uninstall events for participant worker and regular devices, computed over all the days when RacketStore was installed. Workers tend to install apps more often compared to regular users. Concretely, worker devices had an average of 15.94 daily installs (M = 6.41, SD = 27.37) while regular devices had an average of 3.88 daily installs (M = 2.0, SD = 7.29). \newmaterial{KS test and ANOVA reported statistically significant differences between the two groups ($p$-value$<0.05$).}

We recorded fewer daily uninstalls, suggesting that participant devices tend to retain apps: worker devices recorded an average of 7.02 daily uninstalls (M = 2.73, SD = 15.69) and regular devices had an average of 3.29 daily uninstalls (M = 1.8, SD = 6.87). \newmaterial{The KS and ANOVA tests reported significant differences at $p$-value<0.05.} We observe however that several worker devices have a low daily app churn, while some regular devices have a higher daily app churn, making them harder to distinguish based on this feature alone. \newmaterial{We note that the differences in app churn between workers and regular participants could also be due to background differences: ASO workers may be more technically skilled due to the nature of their work. However, we were also surprised by the relatively high app churn of regular users. One reason for this may be that active Instagram users may be more technically skilled than regular people. We did not investigate the technical expertise of participants in our study.}

\begin{figure}
\vspace{10pt}
\includegraphics[width=0.95\columnwidth]{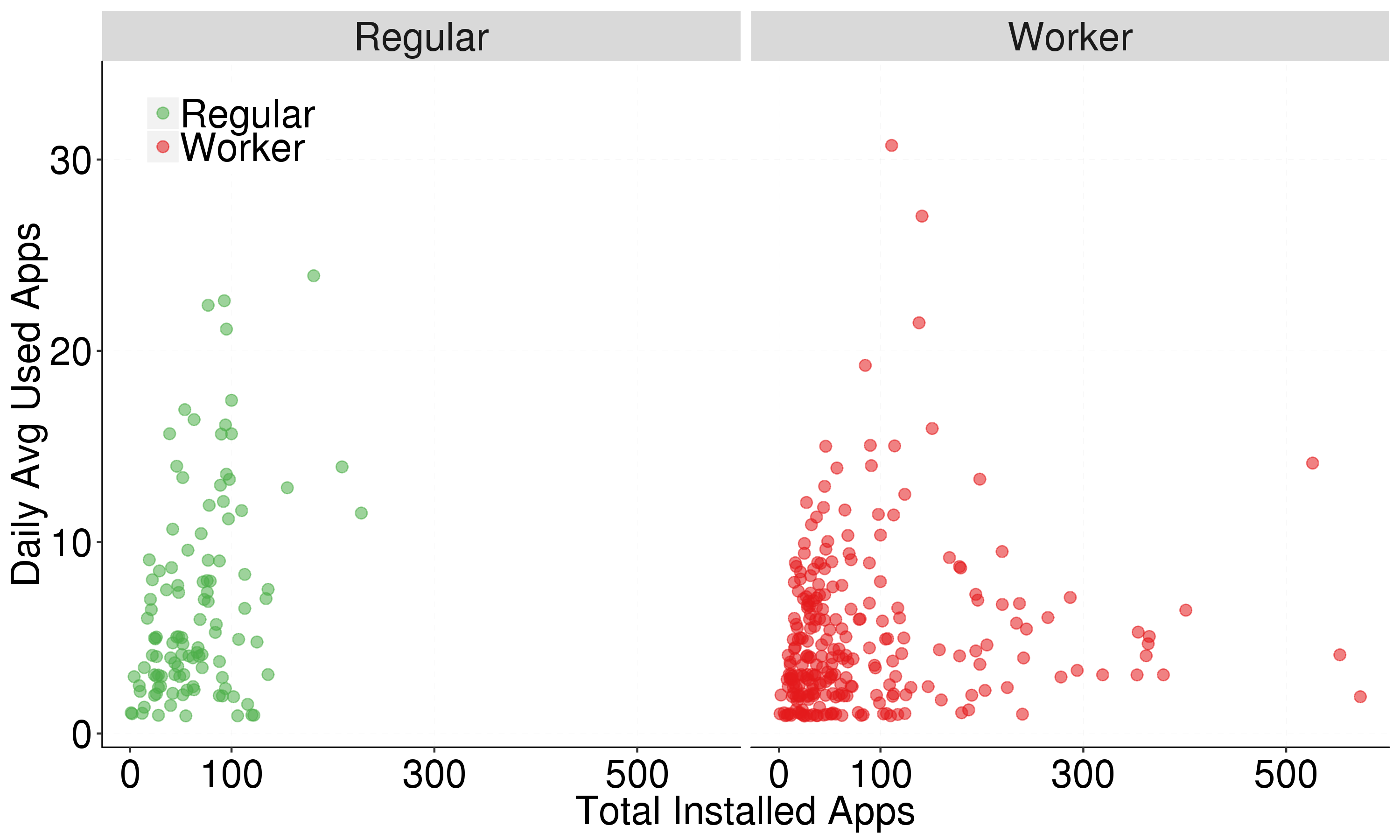}
\vspace{-10pt}
\caption{Scatterplot of the average number of apps used per day per device and the number of apps installed in a device, for regular and worker devices. We observe substantial overlap between regular and worker devices.}
\label{fig:used:installed}
\vspace{-15pt}
\end{figure}

\noindent
{\bf Number of Apps Used Per Day}.
Figure~\ref{fig:used:installed} shows for each of the 141 regular and 399 worker devices in our studies (total of 540), the average number of apps opened per day on the device vs. the total number of apps installed on that device. We observe that several worker devices have many more apps installed than regular devices, and also have more apps used per day. Nevertheless, we also observe substantial overlap in these features between regular and worker devices, perhaps due to the fact that several of the worker devices are organic. This suggests that the daily number of used apps cannot accurately distinguish between worker and regular devices.

\begin{figure}
\vspace{10pt}
\includegraphics[width=0.95\columnwidth]{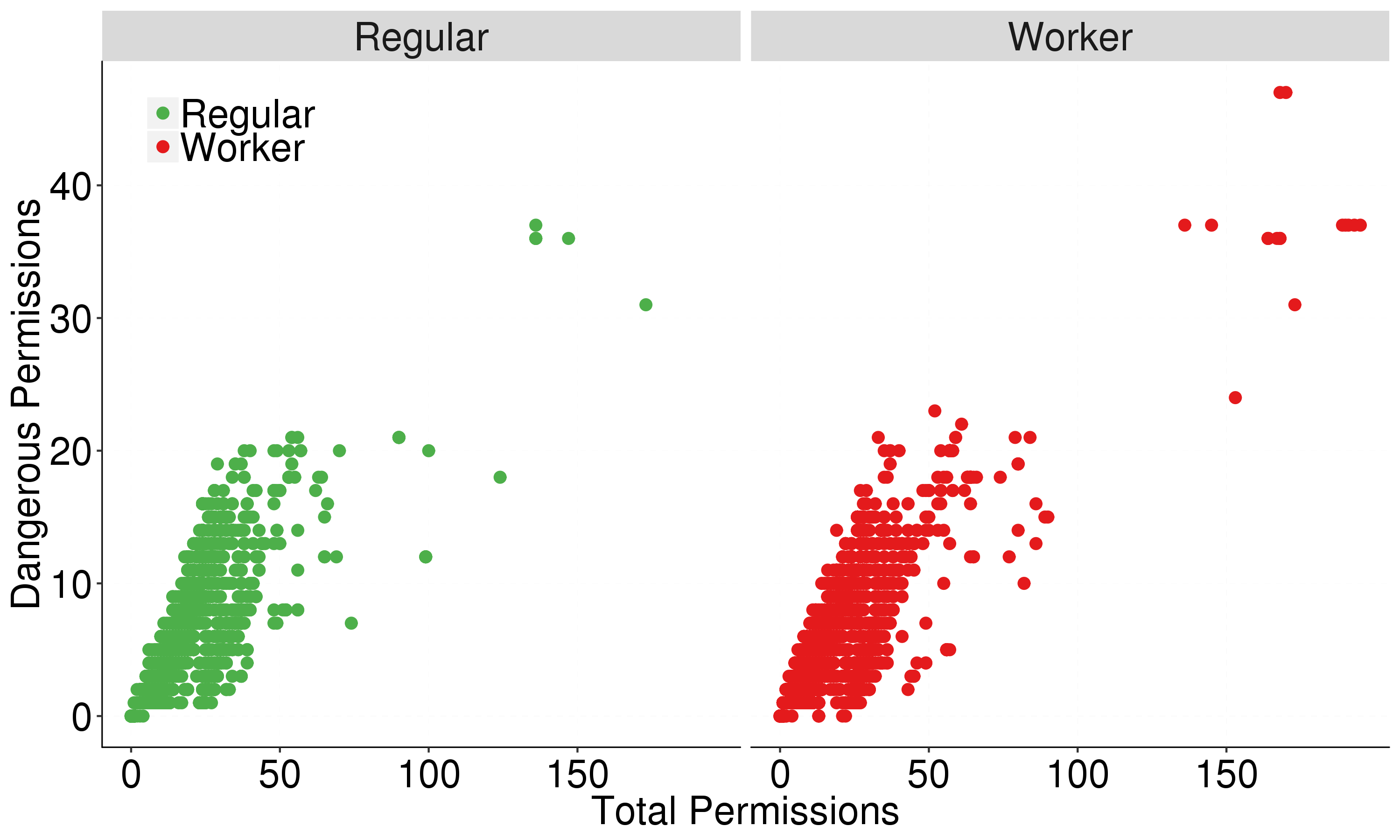}
\vspace{-10pt}
\caption{Comparison of exclusive app permissions for regular and worker devices. Worker devices host apps with the largest ratio of dangerous to total number of permissions.}
\label{fig:permissions}
\vspace{-15pt}
\end{figure}

\noindent
{\bf App Permissions}.
We studied the distribution of permission requirements for unique apps found on participant devices. Figure~\ref{fig:permissions} shows the number of dangerous permissions vs. the total number of permissions for each app found exclusively on regular and worker devices. We found that while some worker devices host apps with the largest number of dangerous permissions, most installed apps share a similar permission profile across all device types. This suggests that the number of permissions requested by an app, including the dangerous permissions, will be ineffective in detecting promoted apps.

We contacted workers to ask about their policy for granting permissions to apps they promote. Five claimed to grant all permissions requested by the apps that they install. However, one participant claimed selective granting, i.e., ``{\it Permissions are given based on the client request. If the client does not ask, we do not give all permissions}''. Four other workers said that there are permissions that they grant grudgingly, e.g., one claimed ``{\it I don't like the location permission because it violates my privacy}'', while two others said claimed to dislike permissions associated with personal data. Two regular participants claimed not to grant all requested permissions. One claimed to avoid granting location permissions, the other was concerned about contacts, images and phone storage permissions.

\noindent
{\bf Summary of Findings}.
In our study, workers posted reviews for a significantly higher number of installed apps. Following app installation, participant workers waited significantly shorter times to post reviews than regular users. While workers tend to install more apps per day than regular users, their devices also had significantly more stopped apps. We conjecture that this happens due to retention install requirements: workers need to keep promoted apps installed, but want to avoid the clutter.

\subsection{Investigation of Malware and Attitudes}
\label{sec:findings:virus}

\begin{figure}
\vspace{10pt}
\includegraphics[width=0.89\columnwidth]{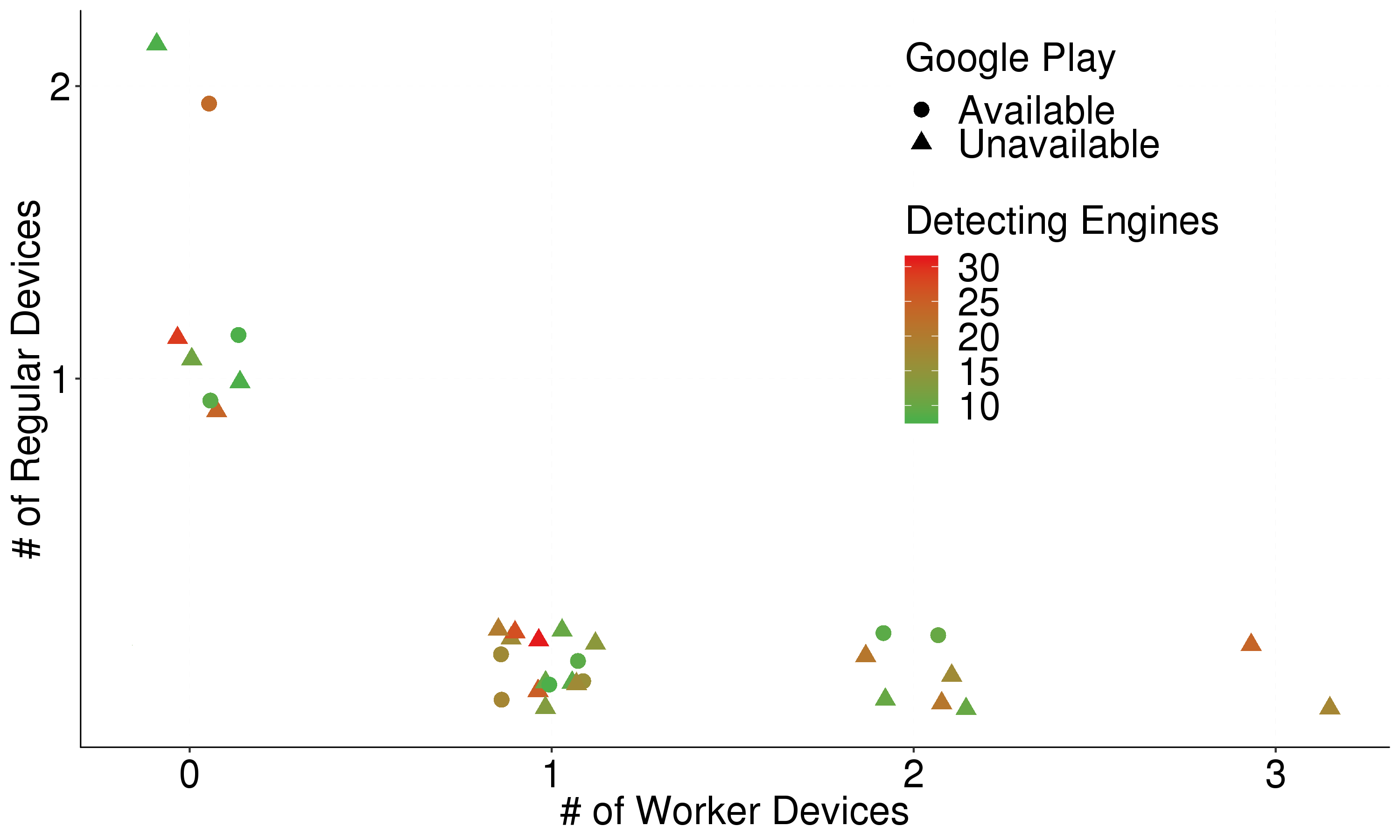}
\vspace{-5pt}
\caption{Comparison of malware occurrence in regular versus worker devices. Each point corresponds to a unique app apk hash, with at least 7 VirusTotal flags. \newmaterial{The color in the legend refers to the number of VT engines that flagged the app.} Worker devices host more unique malware which tends to be present on more devices than for regular users.}
\label{fig:maliciousApps}
\vspace{-10pt}
\end{figure}

We now investigate the potential and perceived impact of malware installations on the workers who participated in our study. We used the VirusTotal research license reports~\cite{total2012virustotal} to analyze the presence of malware on the participant devices. \newmaterial{VirusTotal uses 62 detection engines to process apk files.} We used the \textit{snapshot collector module} ($\S$\ref{sec:racketstore}) to collect 18,079 distinct hashes corresponding to 9,911 unique mobile app identifiers installed in 713 participant devices (549 devices of workers and 164 devices of regular users). The remaining 90 devices did not provide hash information either because of permissions or API incompatibility problems. We collected reports for these hashes in VirusTotal; 12,431 hashes were available in VirusTotal. \newmaterial{We did not collect details about the specific malware families and types detected by the VirusTotal engines, including information about potentially unwanted programs (PUP).}

177 of these apps were flagged malicious by more than one VirusTotal AV tool. We found at least one of these flagged apps in 183 unique devices: 122 devices controlled by workers and 61 devices of regular users. We found 70 unique mobile app identifiers with at least one VT engine flag, that received at least one review from our participants: 64 of these apps were reviewed by workers, and 9 apps were reviewed by regular users.

\newmaterial{Since a single VT flag may be a false positive,} we further compared the occurrence of the most malicious malware samples (flagged by more than 7 VirusTotal engines) in regular user devices versus worker devices. The 7 flag threshold exceeds the value 4 identified in~\cite{PPJJL19}, and most of these samples were later removed from Google Play. Figure~\ref{fig:maliciousApps} shows that malicious samples are more likely to appear in several worker devices when compared to regular users. 

%We observe a couple of apps (ggg.tools.anti01 and com.benstokes.pathakschook), with 15 and 11 engine flags respectively, that were not installed by our official participants, i.e., we never sent them a unique participant ID. Then, I recalled that we actually had people that bypassed our security mechanism (the invitation code) somehow and sent snapshots to our DB anyway. The apps are no longer hosted by the Play Store. \bogdan{Ruben, this may weaken our earlier statements that we collected data only from consenting participants. This may be a problem. I propose to remove this, and also erase the points from the figure. Do you agree?}

\noindent
{\bf Anti Virus (AV) Apps}.
\newmaterial{To determine if participants had sufficient security concerns to install anti-virus (AV) apps}, we first identified 250 anti-virus (AV) apps from Google Play, by doing a search on the app category in the website. We have joined these apps against the apps installed in each of the participant devices that sent at least one snapshot. We found only 19 devices that installed 15 AV apps: 8 worker devices, 7 regular user devices, and 4 unknown (i.e., either Google testing our infrastructure or participants who managed to bypass our invitation code).

\noindent
{\bf Participant Feedback}.
We asked participants if they (1) are concerned about installing malware apps on their devices, (2) have anti virus software installed, and (3) are concerned about the privacy of their device data, including contacts, login info, pictures, videos, text messages, location. Two workers were not concerned about malware or privacy leaks and did not have AV apps installed. One worker said ``{\it I am confident on the ability of my phone to prevent any mishap}''.

Five workers reported being concerned about malware; four claimed to use AV apps. One worker claimed that ``{\it I find a lot of apps like this, which contain a lot of viruses.}'' However, he also claimed to not be concerned about privacy leaks because ``{\it I have 5 devices, 3 mobile devices and 2 computers. 2 out of the 3 mobiles, I use for apps testing and review, 1 mobile I use for my personal work.}''

%Regular participant R1 claimed not to be concerned about installing malware or about privacy leaks, but claimed to have an AV app installed and to have previously installed malware apps. R2 and R3 were concerned about malware and privacy leaks, but only R2 had an AV app installed and to have detected malware.

\noindent
{\bf Summary of Findings}.
Worker-controlled devices have lower instantaneous malware infection rate than regular devices. However, malware installed by worker devices tends to be flagged by more anti-virus engines. Workers had mixed attitudes toward privacy and installing malware and AV apps. This suggests potential vulnerabilities and concerns among workers toward keeping apps installed for longer intervals.

\section{Fake Review Detection}
\label{sec:app}

In this section we investigate whether the app usage data collected by RacketStore ($\S$~\ref{sec:data}), can identify apps installed to be promoted, thus detect fake app installs and reviews. For this, we first introduce app usage features ($\S$~\ref{sec:app:features}) then investigate their ability to train supervised models to distinguish promotion-related vs. personal app use ($\S$~\ref{sec:app:classification}).

\subsection{App Usage Features}
\label{sec:app:features}

\newmaterial{
We extracted the following features for each app installed on participant devices: (1) the number of accounts registered on the device that reviewed the app, before RacketStore was installed, while it was installed, and after it was uninstalled, (2) the install-to-review time ($\S$~\ref{sec:findings:installed}), (3) inter-review times, i.e., statistics over the time difference between all consecutive reviews posted for the app from accounts registered on the device, (4) whether app was opened on multiple days, (5) the number of snapshots per day when the app was the on-screen app, (6) the number of snapshots collected per day from device, (7) {\it inner retention}, i.e., the duration over which the app was installed on the device (while RacketStore was installed), whether the app was installed before RacketStore and was still installed when RacketStore was uninstalled, (8) the number of normal and dangerous permissions requested, (9) the number of permissions requested by the app that have been granted and denied by the user, (10) the number of flags raised by VirusTotal AV tools, and (11) the number of times the app was installed and uninstalled while RacketStore was installed.}

\subsection{App Classification}
\label{sec:app:classification}

We use these features and the datasets of $\S$~\ref{sec:data} to train an app classifier that determines if an app has been installed for promotion or personal use.

\noindent
{\bf Training and Validation Datasets}.
We use the 178 worker and 88 regular devices from which we have received at least two days of fast and slow snapshots. For the other devices we lack enough data to extract good features. 

\begin{table}[]
\centering
\begin{tabular}{c|c|c|c}
\toprule
\textbf{ML Algorithm} & \textbf{Precision} & \textbf{Recall} & \textbf{F1} \\ 
\midrule
\textbf{XGB}          & \cellcolor{blue!25} \textbf{99.78\%}                                 & \cellcolor{blue!25} \textbf{99.67\%}                               & \cellcolor{blue!25} \textbf{99.72\%}                          \\ 
\textbf{RF}           & 99.33\%                                 & 99.23\%                              & 99.27\%                          \\ 
\textbf{LR}           & 99.22\%                                 & 99.00\%                                 & 99.11\%                          \\ 
\textbf{KNN}          & 96.88\%                                 & 96.88\%                              & 96.88\%                          \\ 
\textbf{LVQ}          & 90.99\%                                 & 94.54\%                              & 92.73\%                          \\ 
\bottomrule
\end{tabular}
\caption{Precision, recall, and F-1 measure of app usage classifier (CV $k=10$) using Extreme Gradient Boosting (XGB), Random Forrest (RF), Logistic Regression (LR), K-Nearest Neighbors (KNN), and Learning Vector Quantization (LVQ). XGB performed the best.}
\label{tab:app:classifier}
\vspace{-25pt}
\end{table}

We have set aside randomly selected 20\% (i.e., 38) of the worker-controlled devices and 42\% (i.e., 37) of the regular devices. We use these devices to select a set of train-and-validate apps with suspicious and regular usage. Specifically, we label an app to be suspicious if (1) it was advertised by workers for promotion on the Facebook groups we infiltrated ($\S$~\ref{sec:model}), (2) it was installed on at least five worker devices, and (3) was not installed on any regular devices. The rationale for this selection is that co-installing apps that are not popular and we know have been promoted, is likely the result of ASO work. Further, we label an app to be non-suspicious or regular, if (1) it was not installed on any worker-controlled device, (2) was installed in at least one regular device, and (3) has received at least 15,000 reviews. We have identified 1,041 suspicious apps among the ones installed on the above 38 worker-controlled devices, and 474 non-suspicious apps among those installed on the 37 training regular devices. 

We use these apps to build a {\it train-and-validate app usage dataset} consisting of 2,994 suspicious instances and 345 non-suspicious instances. An instance consists of an app $A$ and a device $D$ on which $A$ has been installed, features extracted from the use of $A$ on the device $D$ ($\S$~\ref{sec:app:features}), and a class label (1 for promotion and 0 for personal usage instance).

\noindent
{\bf Classifier Performance}.
We evaluate the performance of supervised learning algorithms trained with the features introduced in $\S$~\ref{sec:app:features} on the train-and-validate app usage dataset. For this, we use repeated 10-fold cross-validation ($n$=5) over the 2,994 suspicious app usage instances and 345 regular usage instances. Table~\ref{tab:app:classifier} shows the precision, recall and F1-measure of tested algorithms. Extreme Gradient Boosting (XGB) outperforms the other algorithms, achieving an F1-measure of 99.72\%. \newmaterial{KNN achieved best performance for $K=5$.}

\begin{figure}
\includegraphics[width=0.89\columnwidth]{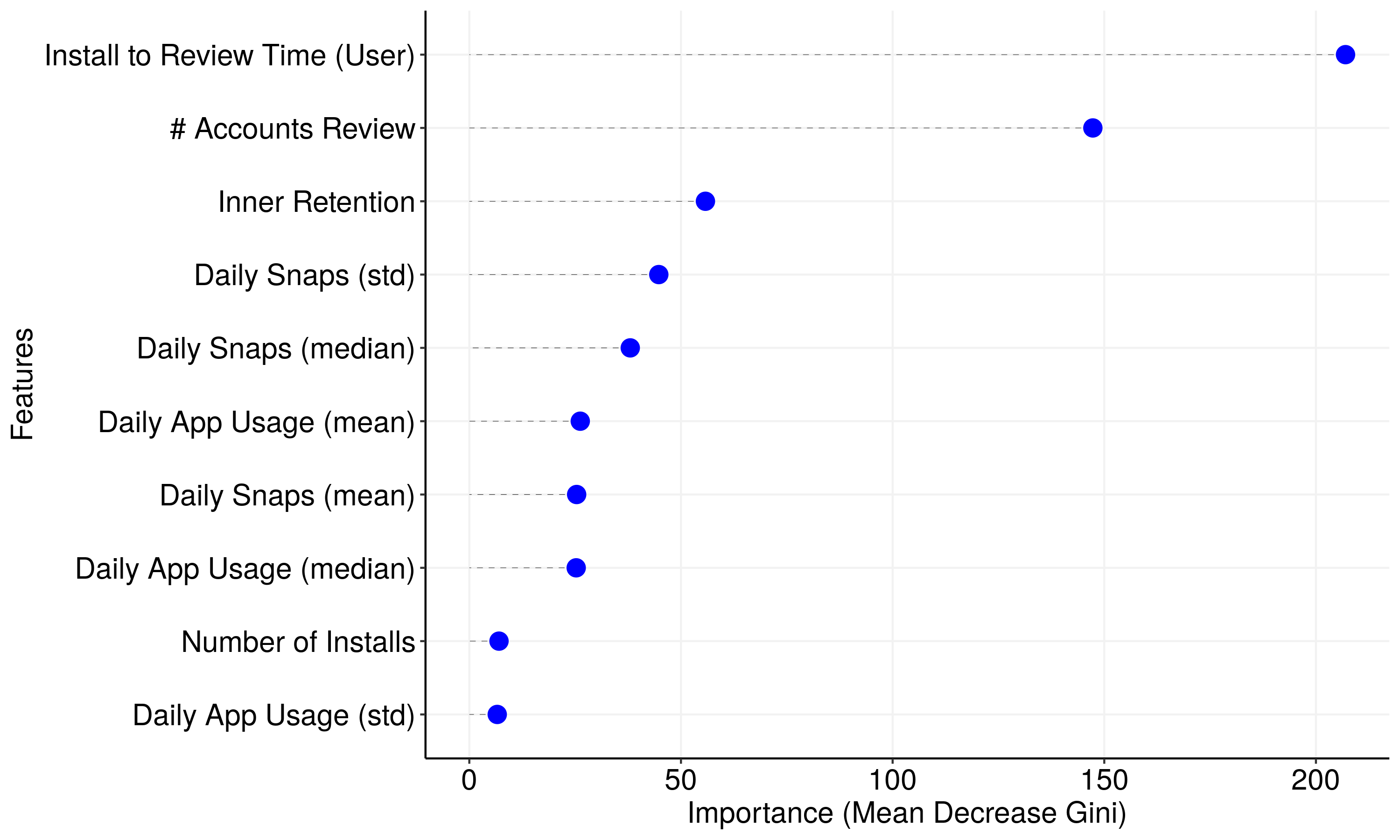}
\vspace{-5pt}
\caption{Top 10 most important features for the app classifier, measured by mean decrease in Gini. The number of accounts that have reviewed the app from the device and the average time between install and review are most important.}
\label{fig:features}
\vspace{-15pt}
\end{figure}

Figure~\ref{fig:features} shows the top 10 most important features to classify app usage, measured by the mean decrease in Gini~\cite{Breiman01}. A higher decrease in Gini indicates higher variable importance. We observe the importance of the number of accounts registered on the device that have reviewed the app ($\S$\ref{sec:findings:accounts}) and the average time between install and review ($\S$\ref{sec:findings:installed}).

\noindent
{\bf Performance Under Balanced Datasets}.
We also evaluate these algorithms trained with balanced datasets of promotion and personal app use instances~\cite{XGCKJLSPFP21}.
Experiments with undersampling the majority class and oversampling the minority class obtain similar performance (F-1 value of 98.76\% and 99.22\% respectively for XGB). The AUC value is over 0.99 across all the algorithms except for KNN where the AUC decreased to 0.90 and 0.92 in undersampling and oversampling respectively. For XGB, the false positive rate is 1.94\% when using oversampling.

\section{ASO Device Detection}
\label{sec:device}

We now investigate whether the device usage data collected by RacketStore ($\S$~\ref{sec:data}) can be used to identify devices controlled by app search optimization workers.

\subsection{Device Usage Features}
\label{sec:device:features}

\newmaterial{
We introduce the following features that model the use of a device: (1) the number of pre-installed and user-installed apps, (2) {\it app suspiciousness}, i.e., the number of apps that were flagged by the app classifier of $\S$~\ref{sec:app}, over the total number of apps installed on the device, (3) the number of apps that were stopped ($\S$~\ref{sec:findings:installed}), (4) the average number of apps installed and uninstalled per day, (5) the number of device-registered Gmail and non-Gmail accounts, and the number of distinct account types, (6) the number of apps installed on the device that have been reviewed from accounts registered on the device, and (7) the total number of apps reviewed by accounts registered on the device.} For most features we use both the user-installed apps and the pre-installed apps, since even the use of pre-installed apps like the app store, e-mail, maps, and browser apps can distinguish regular devices from those controlled by workers.

\subsection{Device Classification}
\label{sec:device:classification}

We evaluate the ability of these features to train classifiers that differentiate between devices controlled by workers and regular users. For this, we use the 178 worker devices and the 88 regular devices that have reported snapshots over at least 2 days. We prioritize precision, since a low precision would lead the app market to take wrong actions against many regular devices~\cite{XGCKJLSPFP21}.

\begin{table}[]
\centering
\begin{tabular}{c|c|c|c}
\toprule
\textbf{ML Algorithm} & \textbf{Precision} & \textbf{Recall} & \textbf{F1} \\
\midrule
\textbf{XGB}    &      \cellcolor{blue!25} \textbf{ 96.81\%}            & \cellcolor{blue!25} \textbf{93.81\%}         & \cellcolor{blue!25} \textbf{95.29\%}     \\ 
\textbf{RF}           & 93.95\%            & 96.06\%         & 94.99\%     \\
\textbf{SVM}          & 96.64\%            & 89.03\%         & 92.68\%     \\ 
\textbf{KNN}          & 94.29\%            & 90.58\%         & 92.40\%     \\ 
\textbf{LVQ}          & 96.40\%            & 82.84\%         & 89.11\%     \\ 
\bottomrule
\end{tabular}
\caption{Precision, recall, and F-1 measure of device classifier (CV $k=10$) using Extreme Gradient Boosting (XGB), Random Forrest (RF),  K-Nearest Neighbors (KNN), Learning Vector Quantization (LVQ),and Support Vector Machines (SVM). XGB performed the best.}
\label{tab:device:classifier}
\vspace{-10pt}
\end{table}

Table~\ref{tab:device:classifier} compares the performance of five supervised learning algorithms trained with the device usage features introduced in $\S$~\ref{sec:device:features}. \newmaterial{KNN achieved best performance for $K=5$.} To balance the worker and regular user device classes, we oversampled the minority class using the SMOTE algorithm ~\cite{Chawla_2002}. We use 10-fold cross-validation over the data from the 178 worker and 88 regular devices. Extreme Gradient Boosting (XGB) outperforms the other algorithms, achieving an F1-measure of 95.29\% and AUC of 0.9455. The precision is 96.81\% and the false positive rate is 1.41\%.

When we undersample the majority class, XGB's recall decreases to 92.97\% with an F-1 value of 95.18\% and AUC of 0.9074. When using no sampling strategy the F-1 increases to 96.86\%, at the expense of the AUC (0.9083).

\begin{figure}
\includegraphics[width=0.89\columnwidth]{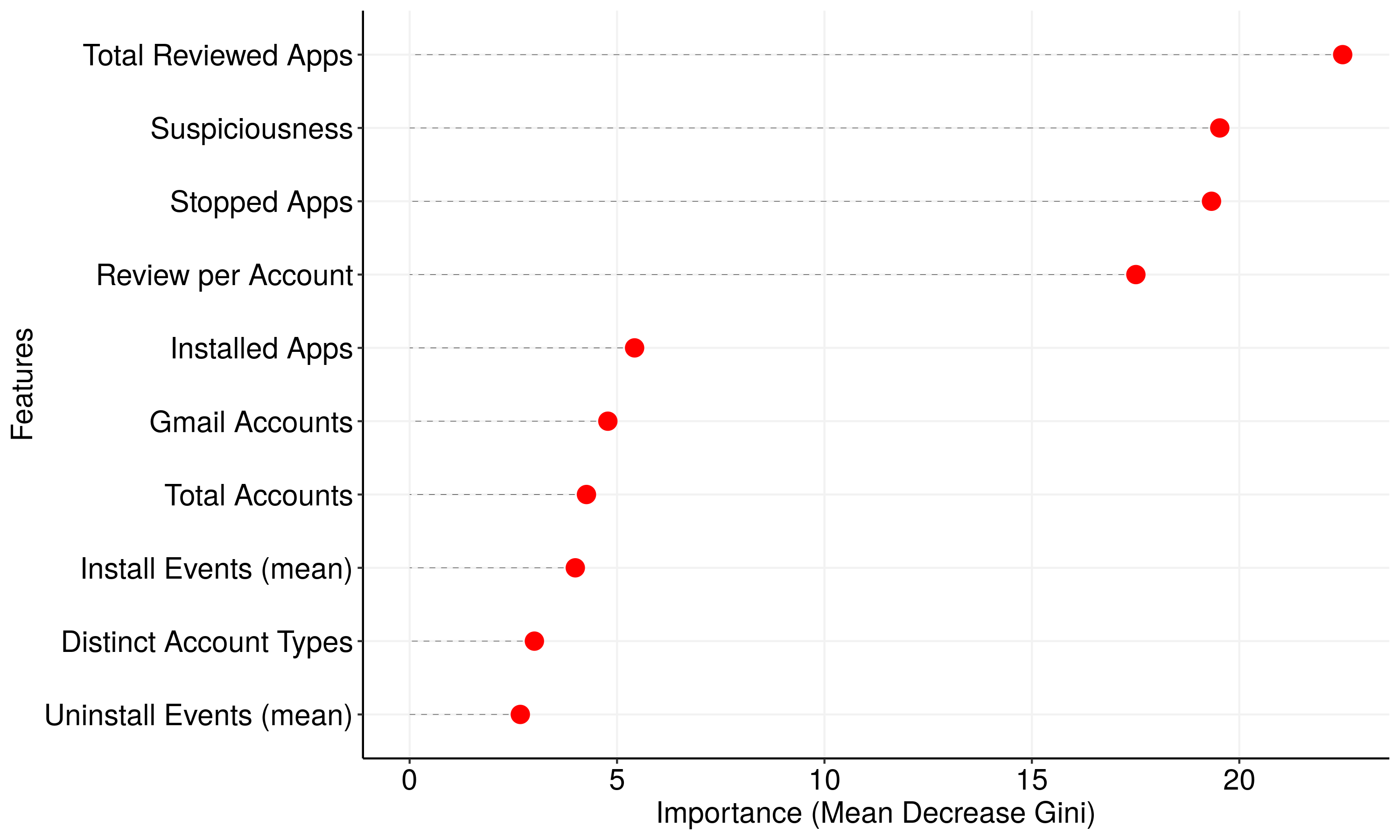}
\vspace{-10pt}
\caption{Top 10 most important features for the device classifier, measured by mean decrease in Gini. This suggests that devices controlled by workers are distinguishable from regular devices on their total number of apps reviewed, percentage of apps used suspiciously, and number of stopped apps.}
\label{fig:device:features}
\vspace{-5pt}
\end{figure}

Figure~\ref{fig:device:features} shows the top 10 most important features in classifying devices as worker-controlled or regular, as measured by the mean decrease in Gini. Four features stand out, confirming their ability to detect worker-controlled devices, i.e., (1) the total number of apps reviewed from accounts registered on the device, (2) the percent of installed apps that were detected to have been used suspiciously by the classifier of $\S$~\ref{sec:app}, (3) the number of stopped apps on the device and (4) the average number of reviews posted from an account registered on the device.

\begin{figure}
\includegraphics[width=0.89\columnwidth]{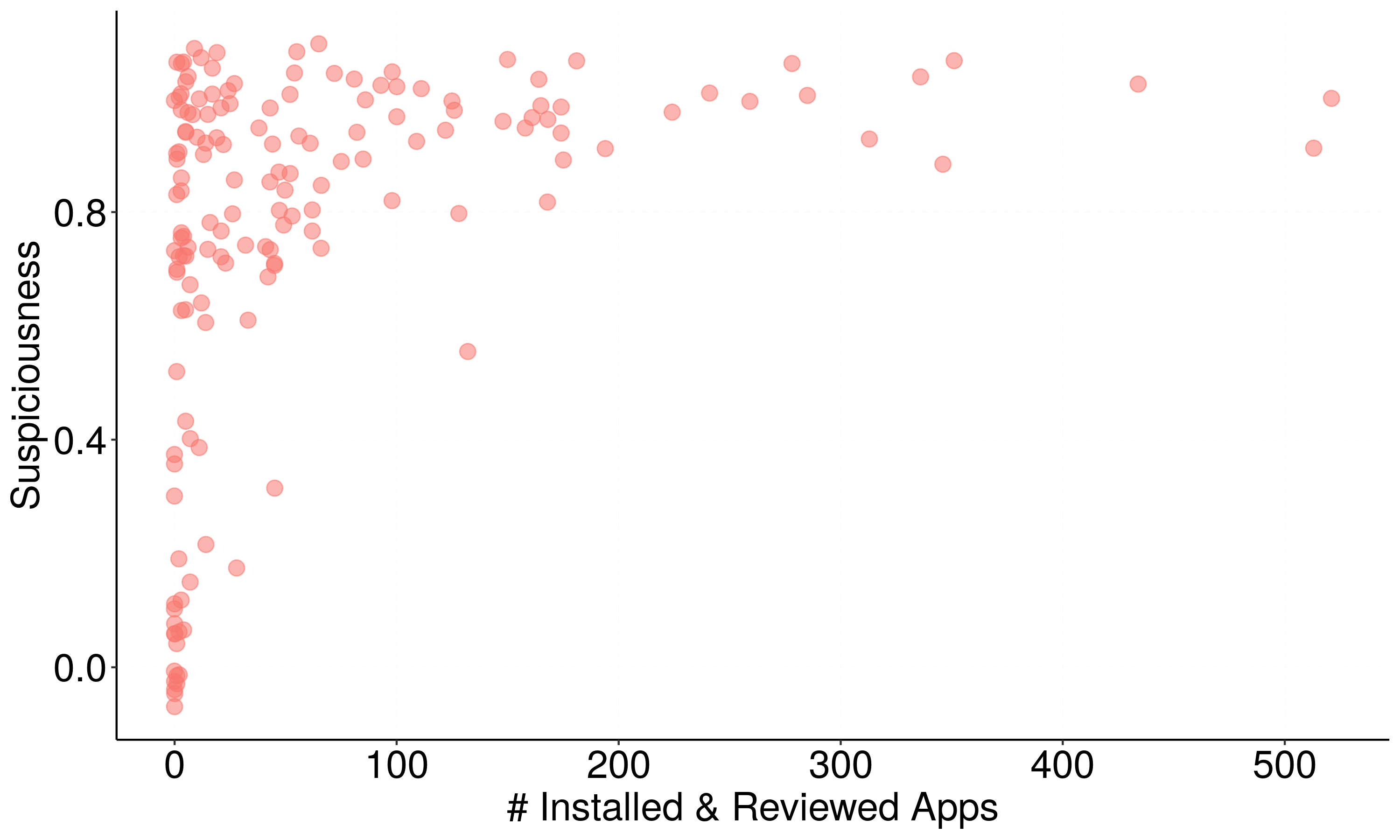}
\vspace{-10pt}
\caption{Scatterplot of 178 worker-controlled devices (one dot per device): app suspiciousness vs. the number of apps installed and reviewed from accounts registered on the device. This reveals that classifiers can detect a range of worker-controlled devices that includes promotion-dedicated devices and devices with organic-indicative usage.}
\label{fig:suspicious:reviews}
\vspace{-15pt}
\end{figure}

Figure~\ref{fig:suspicious:reviews} shows the scatterplot of app suspiciousness vs. the total number of reviewed apps for each of the 178 worker-controlled devices. Out of these 178 devices, 123 devices have organic-indicative behaviors, with at least one of the installed apps being predicted to be used for personal purposes. The remaining 55 devices seem to have been used exclusively for app promotion purposes: all their apps have promotion-indicative behaviors, their median number of Gmail accounts is 31 (M = 37.18, max = 114), and have a median of 23 stopped apps (M = 66.23).

We have manually investigated the devices with high but under 100\% app suspiciousness, and confirmed that such devices are likely to have installed and used apps for personal purposes. Examples include train ticketing apps used at similar times over multiple days, photo gallery apps used in alternation with video players, Samsung pre-installed messaging (\texttt{com.samsung.android.messaging}) and call (\texttt{com.samsung.android.incallui}) apps, and music apps such as Google Play Music being used every day.

However, the classifiers were able to accurately detect even worker-controlled devices with low app suspiciousness, that may belong to novice workers.

\section{Discussion and Limitations}
\label{sec:discussion}

\noindent
\newmaterial{
{\bf Who Should Deploy RacketStore}?
The classifiers proposed in $\S$~\ref{sec:app} and~\ref{sec:device} need more information than what is made publicly available by app stores (e.g., via the Google APIs). Thus, the classifiers can only be effective for RacketStore users. We also note that ASO workers would likely be reluctant to install RacketStore without proper incentives. Therefore, to scale RacketStore's processing of all the apps in an app store, the proposed classifiers should be embedded by app store developers into their pre-installed apps, e.g., the Google Play Store app or Digital Well Being app.} Unlike third-party apps, such clients have by default the permissions to access the required data, and are known to access at least app usage details~\cite{AppUsage,googleWellbeing}.

\noindent
{\bf Privacy-Preserving Classifiers}.
We note that our classifiers need access to sensitive data from user devices, which general users may be reluctant to share. To address this problem, we propose a privacy-preserving approach where our pre-trained models ($\S$~\ref{sec:app} and $\S$~\ref{sec:device}) execute on the user device on locally computed features to detect ASO activities~\cite{tensorflowlite,WXBZCY18}. This approach will only report ASO suspicious activities but no private app and device-usage information. A red flag can be raised if the user uninstalls or blocks this pre-installed client (e.g., the Play Store app), and posts suspicious reviews from accounts registered on such a non-consenting device.

\noindent
{\bf Worker Strategy Evolution}.
ASO workers may attempt to develop strategies to avoid detection by our classifiers. However, our engagement-based features exploit the lack of genuine interest of workers on promoted apps and introduce a tradeoff between detectability and operational costs and exposure to malware. These features include the number of accounts registered on the device, the interval between app install and first review, and user interaction with the app (opened, daily app usage) see $\S$~\ref{sec:app:features}.

\newmaterial{Workers may attempt to manipulate these features using perhaps existing software.} However, workers will still need to keep promoted apps installed for longer intervals, wait more before reviewing them, and interact more with them. When promoted apps are malicious, and workers use personal devices, this may increase the exposure of their personal information to threats. In $\S$~\ref{sec:findings:virus} we show that workers expressed awareness and concern about malware apps. Further, to avoid detection, workers will be forced to register fewer accounts and post fewer reviews for promoted apps from these accounts. This can significantly reduce the amount of fraud posted, thus reduce worker profits from ASO activities.

\noindent
{\bf Recruitment Bias}.
We contacted ASO workers using Facebook groups dedicated to product promotion, and recruited only those who responded, were English speakers, and were willing to participate after approving the consent form. Our Instagram recruitment process reached 61,748 Instagram users who speak English, are of restricted age, show interests related to Android applications, and were willing to participate after approving the consent form.

To reduce the impact of cultural factors in our analysis, we have attempted to recruit both workers and regular users from roughly the same regions. \newmaterial{While the distribution of workers and regular users is not uniform for most countries of our participants, 96\% of the worker devices and 92\% of the regular devices seem to be (according to the unreliable IP-based geolocation) from the geographically close Pakistan, India and Bangladesh.}

\newmaterial{We do not claim that our results generalize to all workers and regular users, including from the same and other countries. Further, workers accessible through other recruitment channels, e.g.,~\cite{TapJoy} may have different behaviors and strategies.} A larger scale recruitment process may identify further types of ASO workers and more diverse regular users. However, the data that we collected from 803 participant devices provides evidence on the ability of device and app usage data to detect the devices controlled, and the reviews posted by different types of workers.

\noindent
\newmaterial{
{\bf Classifier Performance}.
Several machine learning algorithms achieve an F1-measure that exceeds 99\% for the app classification problem ($\S$~\ref{sec:app:classification}), while one algorithm achieved an F1-measure over 95\% for the device classification problem ($\S$~\ref{sec:device:classification}). The investigation in $\S$~\ref{sec:findings} provides an intuition for the ability of several features to help classifiers distinguish between apps and devices used for personal purposes vs. ASO work. This suggests that these algorithms did not overfit the data. Further, the success of these classifiers suggests that standard ML algorithms are suitable and preferable for these classification problems, where they can provide valuable interpretation.}

\newmaterial{
We acknowledge however that the relatively small and biased data that we used to train the app and device classifiers (see recruitment bias above) may lead to reduced applicability to data from other ASO workers and regular users.}

\noindent
{\bf Influence of RacketStore on Participant Behaviors}.\\
Knowledge of being monitored might have influenced participant behaviors. We note however that all participants, including ASO workers and regular users, installed the same version of RacketStore and were provided with the same information before and during the study. Further, our classifiers were able to distinguish between apps and devices used by ASO workers and regular users, even if ASO workers attempted to modify their behaviors during the study.

\section{Related Work}
\label{sec:related}

Farooqi et al.~ \cite{FFLMSV20} studied the market of incentivized app install platforms (IIP) through a honey app that collects the device id, the list of installed apps and events such as opening the app and in-app interaction. We leverage Farooqi et al.~ \cite{FFLMSV20}'s finding of a lack of interest in the app among the workers that installed it for money. RacketStore extends Farooqi et al.~ \cite{FFLMSV20}'s work by collecting and analyzing additional key data that notably includes the list of user accounts registered on the participant device, the reviews posted from those accounts, and the foreground app at 5s intervals. This data enables us to claim a first success in identifying organic ASO activities. Further, our study involved diverse types of ASO workers that we recruited from Facebook groups, and regular users that we recruited using Instagram ads.

Our work is particularly relevant in light of findings that some ASO workers have evolved strategies~\cite{ZXLHZLR18, RHRAC19} to evade detection by both app stores and academic solutions, e.g.,~\cite{CYYP14, XZ14, SMJEKV15, TZXZZ15, YKA16, XZ15.SIAM, JCBFY14, SLK15, XZ15.WiSec, LFWMS17, YMGYLSWL19, NAGKV19, XWLY12, FMLHCG13, HTS16, LFWMS17, HSBGAKMF16, SLLTZ16, XZLW16, X13, GGF14, BXGPF13, LNJLL10, MVLG13, MKLWHCG13, KCS18, LCNK17, YKA16, FLCS15, DFJKS14, CYYP14, LNJLL10}. For instance, Zheng et al.~\cite{ZXLHZLR18} report the emergence of organic workers who attempt to mimic the behavior of real users. Rahman et al.~\cite{RHRAC19} provide insights from studied ASO workers, that confirm the existence of organic workers in the wild. In this paper we provide measurements from devices of ASO workers and regular Android users. Our data suggests that the use of apps installed for promotion differs from that of apps used for personal purposes. Further, even organic workers tend to use their devices in a manner that distinguishes them from regular users.

Related efforts also include extensive work to detect malware Android apps, e.g.,~\cite{OMACRS19, MRGFCKV18, NBF19, BCIFKV15, RAMB16, ZWZJ12, GZZZJ12, RFWOVE19, SB20}. Notably, Yang et al.~\cite{YXALXE15} differentiate malware from benign apps based on the contexts that trigger security-sensitive behaviors. RacketStore detects ASO-promoted apps and devices of workers based on the context of the user interaction with them. While we seek to detect worker interactions with apps, we note that ASO work has been shown to be used to promote malware apps and improve their search rank, thus increase their consumer appeal~\cite{RRCC17}.

Our study of the fraud market for Google services is related to other exploration of fraud markets~\cite{MDSVT19, TGSP11, TMGKP13, SWEKVZZ13}. For instance, Dou et al.~\cite{DLLDLY19} developed a honeypot app and collect data to detect fraudulent bot-generated downloads. Mirian et al.~\cite{MDSVT19} explore the market for Gmail account hijacking by creating synthetic but realistic victim personas and hiring services to hack into such accounts, while DeBlasio et al.~\cite{DBGVS17} characterize the search engine fraud ecosystem using ground truth data internal to the Bing search engine. Stringhini et al.~\cite{SWEKVZZ13} studied Twitter follower markets by purchasing followers from different merchants and used such ground truth to discover patterns and detect market-controlled accounts in the wild. In this paper we leverage our finding of an abundant fraud market for Google services (i.e., review groups with tens of thousands of members) to recruit hundreds of worker-controlled devices, study their usage, and propose solutions to detect and distinguish them from devices used for personal purposes.

\section{Conclusions}
\label{sec:conclusions}

In this paper we have developed RacketStore, the first platform to collect detailed app and device usage information from the devices of app search optimization workers and regular users of Google Play services. We have presented empirical data from RacketStore installs on\FPeval{\result}{clip(\numregular+\numworkers+\numexchangers)} \numprint{\result} devices and from interviews with some of their owners. We have developed a classifier to identify apps installed solely to be promoted and we have shown that on our data, it achieves an F1-measure that exceeds 99\%. We have shown that features that model the user interaction with a device can be used to detect even organic devices with low levels of ASO work hidden among personal activities. Our techniques are resilient to worker strategy modifications, that would impose high overhead on the operation of their devices and the usage of the apps that they promote.

\section{Acknowledgments}

This research was supported by NSF grants CNS-2013671 and CNS-2114911, and CRDF grant G-202105-67826. This publication is based on work supported by a grant from the U.S. Civilian Research \& Development Foundation (CRDF Global). Any opinions, findings and conclusions or recommendations expressed in this material are those of the author(s) and do not necessarily reflect the views of CRDF Global.\\\\\\\\\\

\bibliographystyle{ACM-Reference-Format}
\bibliography{anonymous,attribution,bogdan,crowdsource,ethics,malware,ml,online.fraud,reviews,social.fraud,survey,difprivacy,racketstore}

\appendix

\section{Snapshot Fingerprinting}
\label{appendix:fingerprinting}

To properly analyze the data collected from participating devices, we needed to map each collected device snapshot to a single device. As mentioned in $\S$~\ref{sec:racketstore}, the first snapshot from a device includes (1) the 10-digit {\it install ID} computed by RacketStore upon installation, (2) the 6-digit {\it participant ID}, uniquely generated by us and assigned to each participant, and (3) the Android ID. We expected that the combination of the participant ID, install ID and Android ID will be enough to provide this mapping.

However, we found that the same device can be responsible for multiple install events of RacketStore, i.e., where a different combination of install ID, participant ID and Android ID is reported in different snapshots from the same device. For instance, we encountered cases of different ASO workers, with different assigned participant ids, who shared some devices. This can occur for instance if the workers are employed by the same organization, thus have access to a common set of devices or (2) the same workers pretending to be a different worker. Such workers can install RacketStore at different times believing this to be a repeat campaign. We also observed workers who repeatedly install and uninstall RacketStore, in order to get paid multiple times for the installation. Further, for some installs, due to suspected incompatibilities (there are over 24,000 types of device models), the collected snapshots did not include the Android ID and device information. We note that we did not collect device IMEI since it requires an additional dangerous permission which we wanted to avoid.

To address this problem, we used the following process to fingerprint snapshots. Specifically, we first grouped all the collected device snapshots into $n$ {\it candidate devices}, based on their install ID. We then compared the $\binom{n}{2}$ pairs of candidates to identify and coalesce candidate devices with different install IDs that are actually the same device. First, for each install ID $x$, we compute the RacketStore {\it install interval} $[T_{f}, t_{l}]$ where $t_{f}$ and $t_{l}$ are the first and last timestamp recorded in our database from snapshots that belong to $x$. We then declare as different devices, install pairs $(x,y)$ that have overlapping installation intervals. We then coalesced candidate device pairs that do not overlap on installation intervals based on their Android ID (if present): if the pairs have the same Android ID the two installs belong to the same device, otherwise they are different devices. To validate this approach, we have computed the Jaccard similarity between candidate device pairs, i.e., (1) their sets of tuples $(a,t)$ where $a$ is an app and $t$ is the install time registered by the Android API for app $a$, and (2) their sets of registered accounts. Candidate device pairs with different Android IDs had a Jaccard similarity for installed apps of at most 0.5625. Candidate device pairs with Jaccard similarity above 0.53 for registered accounts, had low similarity for installed apps.

\section{Recruitment Material}
\label{sec:collection:materials}

\begin{figure}
\vspace{10pt}
\includegraphics[width=0.8\columnwidth]{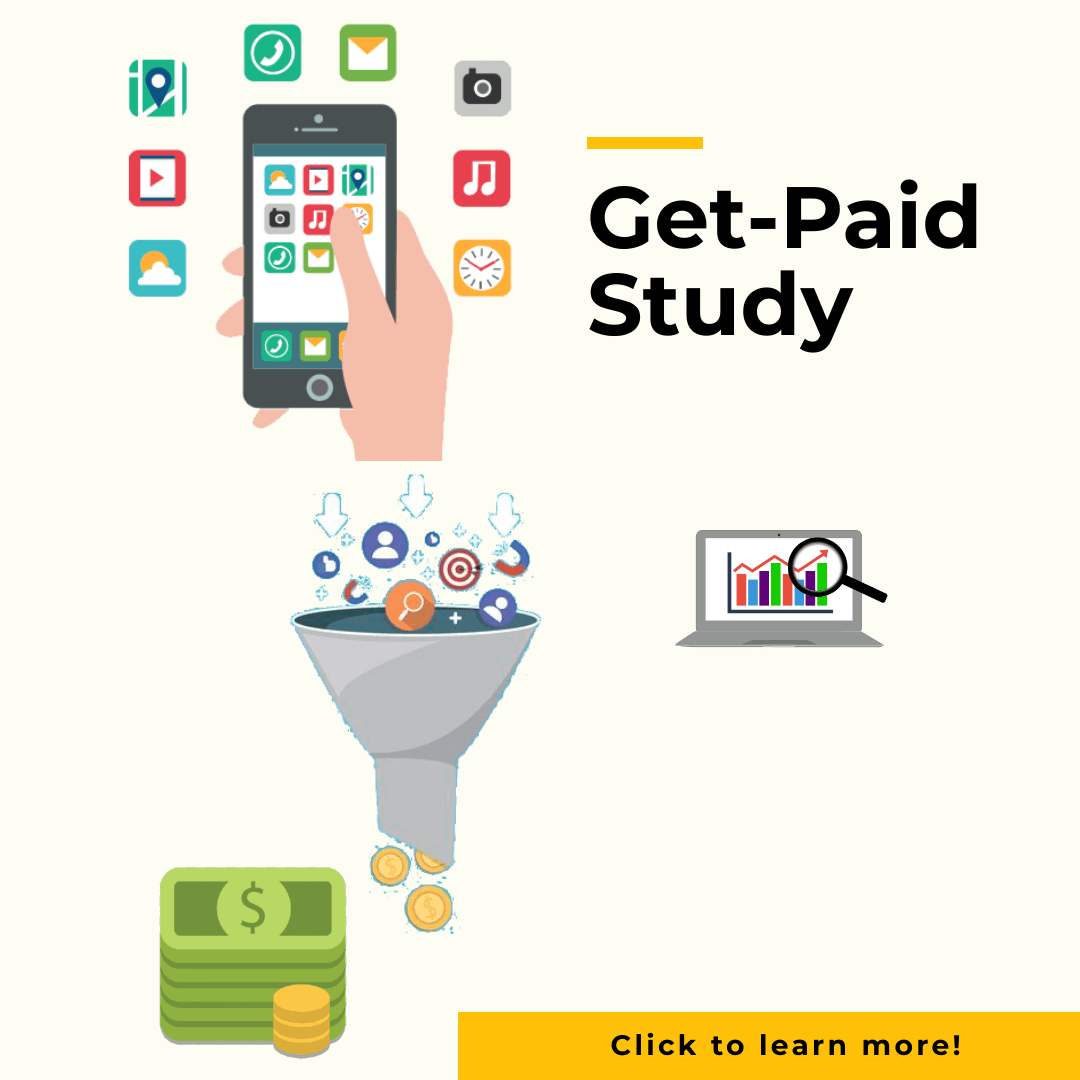}
\vspace{-10pt}
\caption{Ad shown to audience on Instagram Feed, Explore, and Stories. Upon clicking, users are sent to the website where the study is explained, see Figure~\ref{fig:flock:web:screenshot}.}
\label{fig:ad}
\vspace{-15pt}
\end{figure}

\begin{figure}
\centering
\subfigure[]
{\label{fig:landing}{\includegraphics[width=0.89\columnwidth]{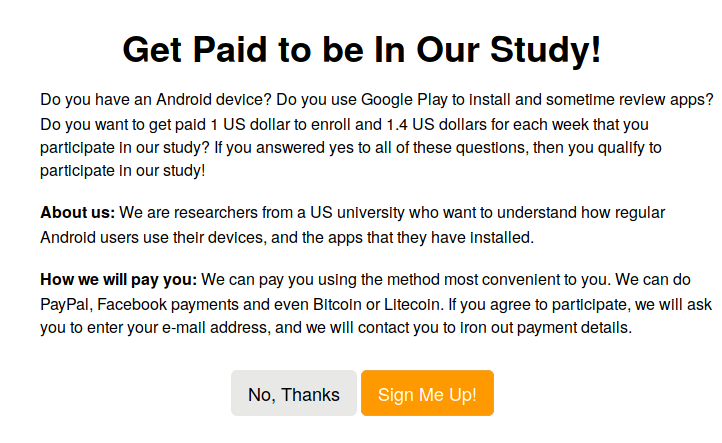}}}
\subfigure[]
{\label{fig:registration}{\includegraphics[width=0.89\columnwidth]{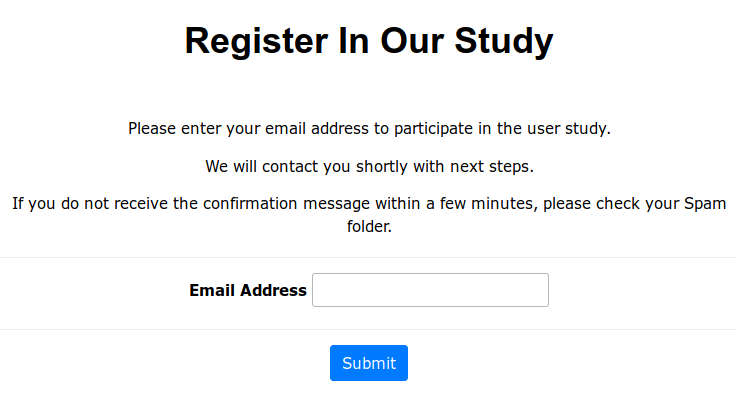}}}
\vspace{-10pt}
\caption{Screenshots of the user study web page. (a) User study landing page. (b) Registration page shown only after user has consented to participate. We ask users to enter their email addresses to contact them with next steps.}
\label{fig:flock:web:screenshot}
\vspace{-15pt}
\end{figure}

Figure~\ref{fig:ad} shows a snapshot of the ad we have shown to audience on Instagram Feed, Explore, and Stories to recruit regular users. Upon clicking, users are sent to a website, shown in Figure~\ref{fig:landing} where we explain the study. Figure~\ref{fig:registration} shows the registration page, displayed to the user only after the user has clicked Sign Me Up and read the content on a web-page that shows the following recruiting message. This message is also shown to recruited ASO workers on a one-on-one basis over messenger apps.

%To recruit participants in our studies, we have used the following recruitment message:

``We are researchers from a US university, looking for people who write paid reviews in Google Play, and are willing to participate in a user study. We are conducting this study as part of an effort to increase our understanding of how app search optimization workers interact with Google Play apps. 
\newline
	If you agree to participate in the study, we will ask you to install an app from Google Play and keep it installed for at least two days. We will pay you \$1 when you install the app. We will then pay you 20 cents for each day when you keep the app installed, on a weekly basis. That is, we will pay you \$1.40 per week, for just keeping the app installed. We may also ask you to use the app to write reviews. If this happens, we will pay you additional money, at a rate that we will negotiate.
\newline
    Please note that we will guard the information that you provide and that we collect, with the utmost secrecy. We will never reveal to anyone any information that may be linked to you, including the fact that you participated in our study
\newline
	Your participation is completely voluntary and you may choose to withdraw at any time. If you agree to participate, please reply to this message. Also, please send us answers to the following questions:
				
	1. Have you ever written paid reviews in Google Play?
				
	2.  How many user accounts do you control in Google Play?
				
	3. How many mobile devices do you own or can access?
				
	4. On how many devices can you install our app?
				
	5. For how many days can you keep our app installed?
				
	6. Are you an administrator or do you post reviews yourself?
				
	7. How many ASO jobs are you currently working on?''

\begin{figure}
\centering
\subfigure[]
{\label{fig:screenshot1}{\includegraphics[width=.49\columnwidth]{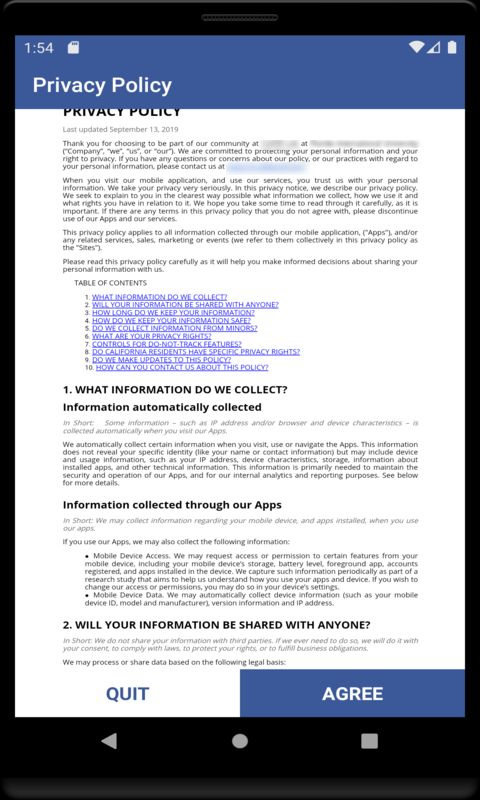}}}
\subfigure[]
{\label{fig:screenshot2}{\includegraphics[width=.49\columnwidth]{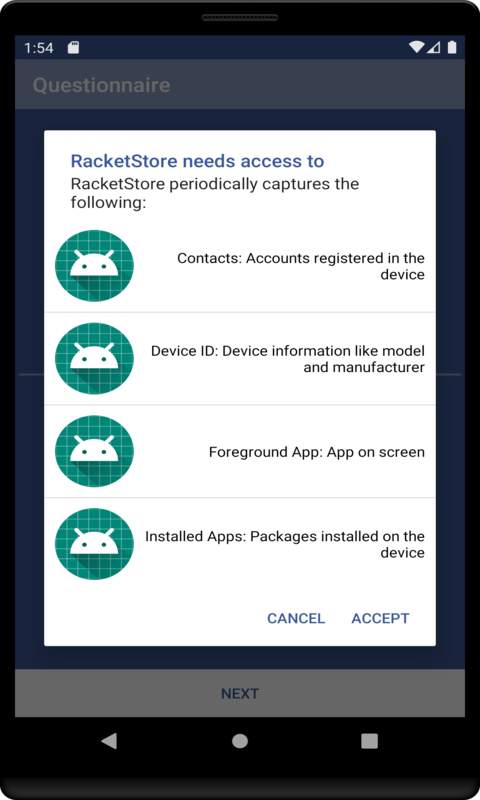}}}
\vspace{-10pt}
\caption{RacketStore mobile app: screenshots of the first two screens shown upon installation. (a) A privacy policy that comprehensively discloses how RacketStore accesses, collects, and uses user data. (b) In-app disclosure to summarize data being collected. RacketStore does not access or collect any personal or sensitive data until the user consents.}
\label{fig:racket:screenshot}
\vspace{-18pt}
\end{figure}

Figure~\ref{fig:racket:screenshot} shows screenshots of information shown to participants by the RacketStore app upon installation. They include the privacy policy that discloses how RacketStore accesses, collects, and uses user data (Figure~\ref{fig:screenshot1}) and the in-app disclosure that summarizes the data collected (Figure~\ref{fig:screenshot2}). %RacketStore does not access or collect any personal or sensitive data without consent.

%Figure~\ref{fig:racket:screenshot} shows screenshots of the RacketStore mobile app, including the privacy policy and in-app disclosure describing the information collected.
	
\section{Consent Form Excerpts}
\label{appendix:consent}

The consent form that we presented to each participant includes the following information:

\paragraph{Procedures}
If you agree to be in the study, we will ask you to do the following things:

1. You will need to click on “I Approve” checkbox at the end of this form.
2. We will send you a link for the RacketStore app in Google Play and you will install it.
3. You will need to keep the app installed as long as you desire to participate in the study, but preferably for at least two days.

Please be assured that your participation in this study is confidential. We will not make your answers public, and we will store them securely (under password protection) in the computers of the researchers.

\paragraph{Risks And/Or Discomforts}
Risks of participating in this study do not exceed those that you would encounter in a regular interaction with other prospective employers. Findings from this study may be used by providers like Google to attempt to detect and eliminate fraud. However, we will never share your data with anyone, for any reason. In fact, we will only store and publish aggregate information from multiple participants that we recruit. Aggregates will not be useful to someone to, e.g., de-anonymize you or any of the participants.

\paragraph{Use of Your Information}
\label{appendix:consent:use}

We automatically collect certain information that includes device and app usage information, your IP address, device characteristics, storage, information about installed and used apps, and other technical information. We also collect public information hosted on the Google Play Store for the reviews that you posted. We collect such information periodically as part of a research study that aims to help us understand how you use your apps and device in order to detect review manipulation or fraud. If you wish to change our access or permissions, you may do so in your device’s settings.

Mobile Device Access. We request access or permission to certain features from your mobile device, including your mobile device’s storage, battery level, foreground app, accounts registered,
and apps installed in the device.

Mobile Device Data. We automatically collect device information (such as your mobile device ID, model and manufacturer) and version information.

\paragraph{Confidentiality}
The records of this study will be kept private and will be protected to the fullest extent provided by law. In any sort of report we might publish, we will not include any information that will make it possible to identify a subject. Research records will be stored securely and only the researcher team will have access to the records. However, your records may be reviewed for audit purposes by authorized University or other agents who will be bound by the same provisions of confidentiality. Those agents will be unable to recover your identity as we do not collect any personally identifiable information.

\paragraph{Compensation}
You will receive a payment of \$1 for installing the app. You will then receive 20 cents for each day when you keep the app installed, on a weekly basis. That is, you will be paid \$1.40 per week, for just keeping the app installed.

\paragraph{Right to Decline or Withdraw}
Your participation in this study is voluntary. You are free to participate in the study or withdraw your consent at any time during the study. Your withdrawal or lack of participation will not affect any benefits to which you are otherwise entitled.

\section{Additional Ethical Considerations}
\label{appendix:ethics}

\noindent
{\bf Privacy Policy}.
We included RacketStore's {\it privacy policy} both in its Google Play profile page, and inside the app, right after the main layout that summarizes the study. The privacy policy explains the data collection process in terms of the information being collected, including that we collect the data periodically and that we collect the reviews posted from the accounts registered on the device (see Figure~\ref{fig:screenshot1} in Appendix~\ref{appendix:consent}). The in-app disclosure (Figure~\ref{fig:screenshot2}) further summarizes the privacy policy. In addition to the consent form, participants also had to give explicit consent to the in-app disclosure. The RacketStore app only collected data if the user provided explicit consent. Therefore, all consenting participants were aware of all the data that we collected. 

\noindent
{\bf Requested Permissions}.
RacketStore explicitly requests two Android permissions, \texttt{PACKAGE\_USAGE\_STATS} and \texttt{GET\_ACCOUNTS}, which participant need to explicitly grant. If any permission is not granted, RacketStore does not collect the corresponding information.

\noindent
{\bf Data Protection}.
We used GDPR~\cite{parliament2016regulation} recommended pseudonymisation for data processing and statistics, and other generally accepted good practices for privacy preservation. \newmaterial{At the completion of the study, we deleted all PII.} We also only generated aggregated statistics. No PII of our participants was disclosed outside of the research team.

\begin{table}[]
\centering
\begin{tabular}{c|c|c|c}
\toprule
\textbf{PII} & \textbf{Collector} & \textbf{Reasons} & \textbf{Deletion} \\ 
\midrule
\textbf{Accounts} & RacketStore & Classification & After use\\
\textbf{Accounts} & RacketStore & Review collection & After use\\
\textbf{Email} & Website & Recruitment & After use\\
\textbf{IP address} & Backend & Statistics & Not stored\\
\textbf{Device ID} & RacketStore & Snap. fingerprint & After use\\
\textbf{Payment Info} & Author & Payment & Not stored\\
\bottomrule
\end{tabular}
\caption{\newmaterial{Personally Identifiable information (PII) that we collected, how we collected it, the reasons, and time of deletion.}}
\label{tab:pii}
\vspace{-25pt}
\end{table}

\noindent
\newmaterial{
{\bf Collected PII}.
Table~\ref{tab:pii} summarizes the PII we collected, how we collected it, the reasons, and how long it was stored. It shows that the PII we collected consists of user accounts registered on participant devices, e-mail address, the device IP address and IDs. The RacketStore app collected the accounts and device IDs; the backend server collected the device IP address and the recruitment website collected the participant e-mail address.}

\newmaterial{
We needed registered accounts for the device classification task. We needed the participant e-mail account in order to contact them with details of the main study and for the follow-up study. We used device IP addresses to help us recruit regular users from the same regions with the ASO worker participants. Further, we needed the device IDs to help us fingerprint snapshots, i.e., group collected snapshots by the device that reported them.}

\newmaterial{
We have deleted all accounts, device IDs and IP addresses after use. We have deleted participant e-mail addresses after the study.}

\noindent
\newmaterial{
{\bf Risk To Benefits Assessment}.
The consent form informed participants on the risks associated with the research. We presented risks in relations to the participant regular interactions. We did not store PII after the completion of the study, and only published aggregate data. We believe that participation risks are reasonable in relation to benefits. We informed participants that benefits for participation include helping us understand and model app search optimization, and also raising their awareness to security and privacy risks that stem from installing malware, granting requested permissions, and account/password management strategies.}

\noindent
\newmaterial{
{\bf ASO Legality and Stigma}.
There are no direct local legal policies to criminalize black hat app search optimization in many countries of the Global South. For example, in Bangladesh, there is not direct law to prevent such activities. The law closest to this issue is a recently passed ICT Act that prohibits the dissemination of incorrect information over the Internet~\cite{Bangladesh.ICT}. However, this law has mostly been applied to control the dissemination of politically motivated, unfounded information over social media (see~\cite{Freedom.House, Bangladesh.Criticism, AHGRD17}, for example). However, ASO work has never been addressed by law enforcing agencies. A similar situation is also present in many other countries in the Global South including India and Pakistan. Hence, the job of our participants was not illegal or unethical according to their own law of the land.}

We also asked several participants questions on the legal aspects of their work, {\it Do you need to be careful about anything? What are your common fear or risks?}. Participants claimed that they are not afraid, for instance, one worker said that ``{\it What we're doing is a legal and right job. So no need to be afraid}''.

\newmaterial{
Moreover, ASO work is not stigmatized in most countries in the Global South. HCI scholars in post-colonial computing argue that many ideas that western scholars hold around how computers are used in non-western contexts are often biased by their own experiences in the West~\cite{IVDPG10}. We argue that the stigma around ASO is a similar case: While in many parts of the West, ASO work might occur as a crime, a job that needs to be hidden, that is not true in countries like Bangladesh, India, Pakistan, or Vietnam. Most local citizens do not understand the technical details of ASO, making it hard for them to judge the work, while in fact, any work with computers and the Internet is considered prestigious in many communities~\cite{PLT09}. These arguments further suggest that the use of deception is not required in studies such as the ones we conducted in this work, when recruiting participants from countries in the Global South.}

\noindent
{\bf Compensation and Professional Security}.
The compensation of the participants was determined according to the fair market rate. We made sure that the rate is not so low that the participants were exploited, and also not so high that they were coerced. There might also be a larger concern about the overall impact of our research on ASO work, in general, that might impact their profession. Previous work explains why studying the ASO workers does not impact their livelihood~\cite{RHRAC19}. Nonetheless, we disclosed and explained this possibility to our participants, and we did not hear any concern from any of them.

\end{document}